\PassOptionsToPackage{pdfpagelabels=false}{hyperref} 
\PassOptionsToPackage{usenames,dvipsnames,svgnames}{xcolor}
\documentclass[letterpaper,twocolumn,10pt]{article}
\usepackage{usenix}

\usepackage{amssymb,amsmath}
\usepackage{graphicx}
\usepackage{dcolumn}
\usepackage{bm}
\usepackage{times}
\usepackage{outlines}
\usepackage{breakurl}
\usepackage{amsthm}
\usepackage{amsfonts}
\usepackage{hyperref}
\usepackage{ulem}
\usepackage{xspace}
\usepackage{paralist}
\usepackage{xcolor}
\usepackage{enumitem}
\usepackage{outlines}
\usepackage{graphicx}
\usepackage{booktabs} 
\usepackage{fnpct}

\newcommand{\sysname}[0]{\textsc{UCCL-EP}\xspace}
\hypersetup{pdfstartview=FitH,pdfpagelayout=SinglePage,linkcolor={green!40!black}}

\newtheoremstyle{mystyle} 
  {.3\topsep}               
  {.3\topsep}               
  {\itshape}              
  {}                      
  {\bfseries}             
  {.}                     
  {.5em}                  
  {}                      

\theoremstyle{mystyle}

\newcommand{\ie}{i.e.\xspace}
\newcommand{\eg}{e.g.\xspace}

\newcommand{\eat}[1]{}
\usepackage{multirow}
\usepackage[font=small,labelfont={small,bf}]{caption}
\usepackage[font=small,labelfont={small,bf}]{subcaption}
\usepackage{pifont}
\usepackage{enumitem}
\usepackage{refcount}
\newcommand{\cmark}{\ding{51}}%
\newcommand{\xmark}{\ding{55}}%
\setlist[itemize]{itemsep=0pt, partopsep=0pt, parsep=1pt, topsep=1pt}
\setlist[enumerate]{itemsep=0pt, partopsep=0pt, parsep=1pt, topsep=1pt}

\usepackage{inconsolata}
\usepackage{minted}
\setminted{
  python3=true,
  fontsize=\footnotesize,
  autogobble,
  xleftmargin=1em,
  numbersep=4pt,
  highlightcolor=lightgray,
}
\setmintedinline{fontsize=\small}
\newmintedfile[pythoncode]{python}{}
\newmintedfile[yamlcode]{yaml}{
    baselinestretch=1.2
}

\usepackage{booktabs} 
\usepackage{siunitx} 
\newcommand{\hide}[1] 
{
\ifthenelse{\boolean{false}}{#1}{}
}

\newcommand{\content}[1] 
{
\ifthenelse{\boolean{false}}{#1}{}
}

\newcommand{\summary}[1] 
{
\ifthenelse{\boolean{false}}{}{#1}
}

\usepackage[T1]{fontenc}
\usepackage{textcomp}
\usepackage{setspace}
\usepackage[title]{appendix}

\usepackage{xcolor}

\usepackage[compact,small]{titlesec}
\titlespacing*{\paragraph}{0pt}{0pt}{*}

\newcommand{\amd}[1]{\noindent{\color{black} {#1}}}

\usepackage{times}
\usepackage{float}
\usepackage{multirow}
\usepackage{array}
\usepackage{etoolbox}
\usepackage{breakurl}
\usepackage{makecell}

\usepackage{booktabs}
\usepackage{amsmath}
\usepackage[linesnumbered,ruled]{algorithm2e}
\usepackage[noend]{algpseudocode}
\usepackage{csquotes}
\usepackage[htt]{hyphenat}
\usepackage{upgreek}
\usepackage{url}

\definecolor{mygreen}{rgb}{0.1,0.5,0.1}
\definecolor{myblue}{rgb}{0.1,0.1,0.7}
\definecolor{mycmtcol}{rgb}{0.5,0.0,0.5}

\usepackage[hyperref]{}
\hypersetup{              
  colorlinks,
  linkcolor=myblue,
  citecolor=mygreen,
  urlcolor=myblue,
}

\newcommand{\pfive}{NV\_EFA3\xspace}
\newcommand{\psix}{NV\_EFA4\xspace}
\newcommand{\nebius}{NV\_IB\xspace}
\newcommand{\lambdalab}{NV\_C2C\_IB\xspace}
\newcommand{\amdib}{AMD\_CX7\xspace}
\newcommand{\amdbrc}{AMD\_BRC\xspace}

\usepackage{epstopdf}
\usepackage{graphicx}
\usepackage[T1]{fontenc}
\usepackage[utf8]{inputenc}
\usepackage{blindtext}

\usepackage{arydshln}
\newcommand\module{\@startsection{paragraph}{4}{\z@}%
                                    {1.25ex \@plus1ex \@minus.2ex}%
                                    {-1em}%
                                    {\normalfont\normalsize\slshape\bfseries}}

\usepackage{listings}

\makeatletter
\usepackage{fancyhdr}

\makeatletter
\renewcommand{\@fnsymbol}[1]{%
  \ifcase#1
  \or \textcolor{black}{\ensuremath{\spadesuit}}
  \else \@arabic{#1}
  \fi}
\makeatother
\renewcommand{\thefootnote}{\fnsymbol{footnote}}

\title{\sysname: Portable Expert-Parallel Communication}

\author{
{\rm Ziming Mao$^{\dagger}$~~~Yihan Zhang$^{\ddagger}$~~~Chihan Cui$^{\S}$~~~Zhen Huang$^{\P}$~~~Kaichao You$^{\clubsuit}$~~~Zhongjie Chen$^{\heartsuit}$} \\
\vspace{0.05in}
{\rm Zhiying Xu\thanks{This work does not relate to the position at Amazon.}\enspace~~~~Zhenyu Gu$^{\P}$~~~Scott Shenker$^{\dagger\diamond}$~~~Costin Raiciu$^{\star}$~~~Yang Zhou$^{\ddagger}$~~~Ion Stoica$^{\dagger}$} \\
$^\dagger$UC Berkeley~~~$^\ddagger$UC Davis~~~$^\S$UW--Madison~~~$^\P$AMD~~~$^\clubsuit$Independent Researcher~~~$^\heartsuit$Tsinghua University\\
$^\spadesuit$Amazon Web Services~~~$^\diamond$ICSI~~~$^\star$Broadcom \& University Politehnica of Bucharest\\
}

\begin{document}

\maketitle
\setcounter{footnote}{0}
\renewcommand{\thefootnote}{\arabic{footnote}}

\begin{abstract}

Mixture-of-Experts (MoE) workloads rely on expert parallelism (EP) to achieve high GPU efficiency. \amd{State-of-the-art EP communication systems such as DeepEP demonstrate strong performance but exhibit poor portability across heterogeneous GPU and NIC platforms.}
The poor portability is rooted in architecture: GPU-initiated token-level RDMA communication requires tight vertical integration between GPUs and NICs, \eg, \textcolor{black}{GPU writes to NIC driver/MMIO interfaces.} 

We present \sysname, a portable EP communication system that delivers DeepEP-level performance across heterogeneous GPU and NIC hardware. 
\sysname replaces GPU-initiated RDMA with a high-throughput GPU-CPU control channel: compact token-routing commands are transferred to multithreaded CPU proxies, which then issue GPUDirect RDMA operations on behalf of GPUs. 
\sysname further emulates various ordering semantics required by specialized EP communication modes using RDMA immediate data, enabling correctness on NICs that lack such ordering, \eg, AWS EFA. We implement \sysname on NVIDIA and AMD GPUs with EFA and Broadcom NICs. 
On EFA, it outperforms the best existing EP solution by up to $2.1\times$ for dispatch and combine throughput. On NVIDIA-only platform, \sysname achieves comparable performance to the original DeepEP.
\sysname also improves token throughput on SGLang by up to 40\% on the NVIDIA+EFA platform, and improves DeepSeek-V3 training throughput over the AMD Primus/Megatron-LM framework by up to 45\% on a 16-node AMD+Broadcom platform.

\end{abstract}
\section{Introduction} 
\label{sec:introduction}

State-of-the-art large language models (LLMs), such as
DeepSeek-V3~\cite{deepseek_v3, deepseek_v3_2},
OpenAI gpt-oss~\cite{gptoss}, 
Google Gemini-3 Pro~\cite{gemini3pro},
and Meta LLaMA 4~\cite{llama4},
are increasingly based on the Mixture-of-Experts (MoE) architecture. 
In a MoE layer, a gating network running on GPUs selects a small subset of experts for each token activation, dispatches the token activation to those experts, and then aggregates their output activations.
Modern MoE models typically instantiate hundreds of experts that specialize in different input patterns, so that only a few experts are active for each token.
This sparsity allows MoE models to achieve accuracy comparable to large dense models while using only a fraction of the per-token inference cost, making them the standard choice for many frontier LLMs.

Training and serving large MoE models require expert parallelism (EP), which places different experts on different GPUs and communicates token activations among GPUs in an all-to-all manner. 
By sparsely sharding experts on different GPUs, EP leaves enough GPU memory for matrix multiplication on extremely large batch sizes (\eg, 4096 in DeepSeek-V3~\cite{deepseek_v3}), thus enabling high GPU resource efficiency. 
Expert-parallel communication plays a pivotal role in the EP efficiency~\cite{deepseek_v3, deepseek_isca}, because token activations are small (\eg, 7KB), dispatch and combine operations are frequent (\eg, selecting 8 experts per token), and routing destinations are only determined at runtime in GPUs (\S\ref{ssec:expert-parallel-communication}). 

\textit{GPU-initiated token-level (fine-grained) communication} (\S\ref{ssec:ep_fine_grained_comm}) is an emerging and key communication pattern for efficient token dispatch and combine at runtime, where DeepEP~\cite{deepep2025} by DeepSeek is the most popular communication system implementing it. 
Different from CPU-initiated bulk-transfer (coarse-grained) communication in NCCL/RCCL~\cite{nccl, rccl}, 
DeepEP leverages the advanced NVIDIA IBGDA (InfiniBand GPUDirect Async)~\cite{ibgda} technique that enables GPUs to directly operate RDMA NICs (network interface controllers) to write out small activations. 
Leveraging GPU-initiated token-level communication, DeepEP implements efficient GPU-side token deduplication during dispatch (avoid sending duplicate tokens to experts on the same node) and hierarchical reduce during combine to achieve superior performance. 
DeepEP has been widely adopted by various training and serving frameworks such as Megatron-LM~\cite{megatron_lm_deepep}, vLLM~\cite{vllm_deepep}, and SGLang~\cite{sglang_deepep} 

\begin{figure}[!t]
    \vspace{0.05in}
    \centering
    \begin{subfigure}[t]{0.48\linewidth}
        \centering
        \includegraphics[width=\linewidth]{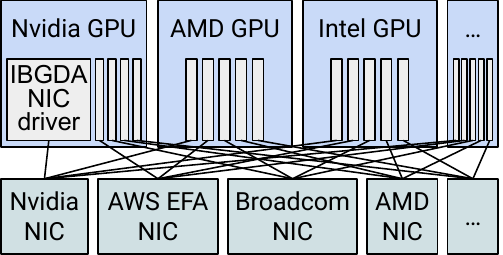}
        \caption{IBGDA-style.}
        \label{fig:ibgda}
    \end{subfigure}
    \hfil
    \begin{subfigure}[t]{0.48\linewidth}
        \centering
        \includegraphics[width=\linewidth]{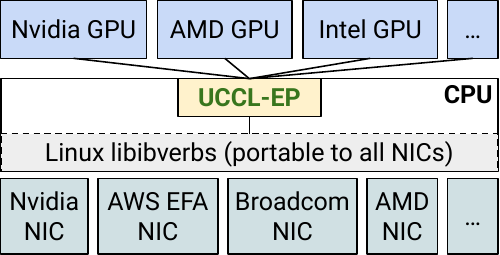}
        \caption{\sysname}
        \label{fig:uep}
    \end{subfigure}
    \vspace{-0.10in}
    \caption{Assuming $m$ GPU vendors and $n$ NIC vendors, \sysname enables $O(m)$ effort, instead of IBGDA's $O(m\times n)$, to support GPU-initiated token-level communication for expert parallelism. 
    }
    \label{fig:intro-figure}\vspace{-0.25in}
\end{figure}

Although GPU-initiated token-level communication leads to high performance, its design unfortunately results in poor portability. There are two key reasons: GPUs directly issuing RDMA operations to NICs prevent interoperability across different GPUs and NICs, as it requires the GPU to \textcolor{black}{write to the NIC's driver-defined MMIO doorbell/register interface}~\cite{mooncake_mlx5_ifc_2025, nvshmem_mlx5_ifc_2025, mori_bnxt_re_hsi_2025}; GPU kernels also impose strict ordering and delivery semantics assumptions on the underlying network, which often mismatch the capabilities and semantics of heterogeneous NICs.
As shown in Figure~\ref{fig:ibgda}, ML infrastructure developers need to vertically integrate the GPU and NIC software ecosystems. This involves complex and subtle code migration and maintenance that are both NIC vendor-specific and GPU vendor-specific. 
Assume there are $m$ types of GPU accelerators and $n$ types of NICs. Developers need to pay $O(m\times n)$ effort to enable such communication on heterogeneous hardware. 
Because of this portability issue, the official DeepEP~\cite{deepep2025} only supports NVIDIA GPUs and NICs, creating severe vendor lock-in and high portability effort for alternative GPU and NIC devices. 
For example, at the time of writing, it remains impossible to run DeepEP on any AWS GPU instances with AWS EFA RDMA NICs; it is also challenging to run DeepEP on any GPUs with Broadcom NICs, one of the major NIC vendors. 
At a high level, such vertical integration is somewhat reminiscent of mainframe servers, which get replaced by less-coupled and more portable commodity servers from heterogeneous vendors in modern cloud computing. 

This paper introduces a new portable expert-parallel communication architecture for GPU-initiated token-level communication with high performance. 
We envision that with our architecture, developers only need to pay $m$ times effort (instead of $m \times n$) to enable systems like DeepEP on heterogeneous GPU and NIC hardware, as shown in Figure~\ref{fig:uep}.

Our key insight is to leverage the host CPU to help break the tight coupling between GPUs and NICs, where the CPU is essentially portable to any GPUs and NICs via the \texttt{libibverbs} library~\cite{libibverbs} maintained by the Linux community and NIC vendors. 
In particular, GPUs and CPUs are connected through high-throughput, low-latency interconnects such as PCIe and even faster NVLink-C2C~\cite{nvidia_gh200_grace_hopper_superchip}, which allows efficient transferring of small control data from GPUs to CPUs. 
By embedding the token dispatching information, such as source and destination address, into the control data, CPUs directly issue GPUDirect RDMA to write out the activations on behalf of GPUs. 
We note that modern GPU servers usually have hundreds of CPU cores, which are often heavily underutilized, \eg, 20\%-45\% CPU utilization reported by several industry companies in their GPU clusters~\cite{characterization_llm_cluster, byteps}; 
From discussion with a model training team inside a major GPU vendor, their model training using Megatron-LM~\cite{megatronlm} yields on average 14.5\% CPU utilization. 

Moreover, flexible host CPUs can help bridge the semantic gap between the guarantees required by specialized EP communication systems and the primitives actually provided by heterogeneous NICs.
For example, DeepEP requires a specific write-then-atomic ordering requirement to inform remote GPUs of data arrival, but such a requirement is typically not enforced by all RDMA NICs, such as AWS EFA NICs, which do not guarantee ordering.

To realize these insights, we need to address two challenges: i) how to efficiently transfer control data from GPUs to CPUs so that the CPU does not become the bottleneck. This is challenging, especially given the frequent token dispatch/combine operations as high as 7Mop/s/GPU under 7KB activations~\cite{deepseek_v3} and a 400G network. ii) How to use CPUs to express and enforce various delivery semantics (\eg, write-then-atomic) in CPU to accommodate heterogeneous NICs that do not support them on their own. 

We present \sysname, a portable expert-parallel communication system with GPU-initiated token-level communication. 
\sysname decouples communication initiation from communication execution: it keeps GPUs initiating communication for fine-grained token control, and delegates the communication tasks to the host CPU. 
\sysname addresses the first challenge with an efficient multithreaded, lock-free FIFO communication channel between GPUs and CPUs. This channel minimizes the PCIe traversing overhead and allows multiple CPU proxy threads to forward and execute the GPU-generated token routing decisions. 
It leverages GPU-side caching and space-efficient message packing that minimizes PCIe operations to achieve millions of messages per second from GPUs to CPUs. 
To address the second challenge, \sysname leverages the immediate data feature that is widely available in RDMA NICs (this immediate data field has been standardized into the RoCEv2 packet header~\cite{ibta_spec}). \sysname embeds sequence information into this immediate data and lets the receiver hold atomic updates until all previous corresponding writes have finished. We show how host CPUs can flexibly bridge the delivery guarantees that GPUs require and the ones actually provided by the networking layer.

We have implemented \sysname and enabled it on various heterogeneous platforms, including NVIDIA GPUs + AWS EFA NICs, and AMD GPUs + Broadcom NICs. 
On the NVIDIA+EFA platform, compared to the second-best EP solution PPLX~\cite{pplx_garden}, \sysname achieved up to $2.1\times$ higher throughput. On the AMD+Broadcom platform, \sysname achieves comparable performance to the original DeepEP on the NVIDIA-only platform. \sysname supports drop-in replacement for any DeepEP applications without any line of code change. It speeds up serving throughput in SGLang~\cite{zheng2024sglang} by up to 40\% on the NVIDIA+EFA platform, and improves DeepSeek-V3~\footnote{We downscale DeepSeek-V3 to 32 layers and 379B parameters to fit onto the 16-node, 128 MI300X GPUs (\S\ref{ssec:eval-training}). } training throughput in the AMD Primus/Megatron-LM framework by up to 45\% on a 16-node AMD+Broadcom platform. 
To the best of our knowledge, \sysname is the first work that enables running GPU-initiated token-level communication on platforms with non-NVIDIA NICs. \sysname is open-sourced\footnote{https://github.com/uccl-project/uccl/tree/main/ep}.

\section{Background}

\subsection{Expert-parallel communication}
\label{ssec:expert-parallel-communication}
Mixture-of-Experts (MoE) architectures have emerged as a leading architectural pattern for building state-of-the-art LLMs as they provide massive parameter capacity while keeping per-token compute low by activating only a small subset of experts. This sparsity enables specialization—experts learn domain-specific behaviors—while preserving general-purpose performance. As a result, many frontier models now adopt MoE designs, including 
OpenAI's gpt-oss~\cite{gptoss}, Google's Gemini-3 Pro~\cite{gemini3pro} and GLaM~\cite{du2022glam}, Mistral's Mixtral 8×7B~\cite{jiang2024mixtral}, DeepSeek's DeepSeekMoE~\cite{dai2024deepseekmoe} and DeepSeek-V3~\cite{deepseek_v3, deepseek_v3_2} family, the Allen Institute's OLMoE~\cite{muennighoff2024olmoe}, and Meta's Llama 4~\cite{meta_llama4_2025}, all of which use MoE to push model capacity and quality without the prohibitive compute costs of dense models. These systems power a wide range of applications—from interactive agents and large-scale pretraining to multimodal reasoning and domain-specialized tasks. 

\begin{figure}[!t]
    \centering
    \includegraphics[width=\linewidth]{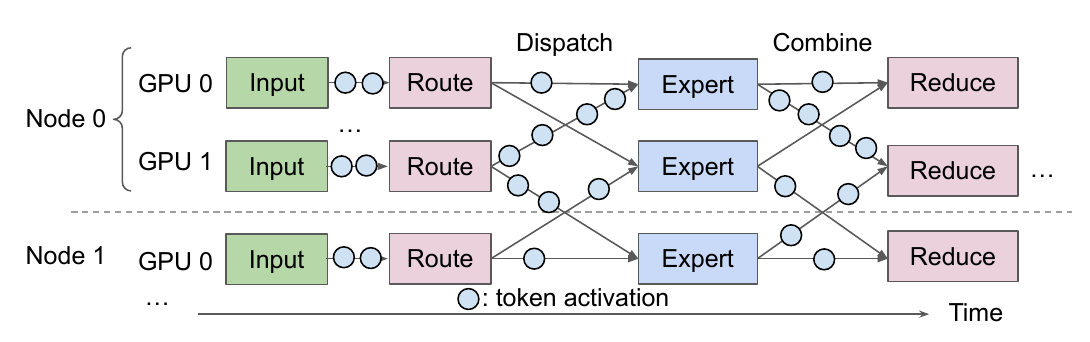}
    \vspace{-0.10in}
    \caption{The MoE communication pattern consists of the dispatch phase and the combine phase. In the dispatch phase, the router sends each token’s input activations to one or more selected experts. In the combine phase, expert activations are collected and aggregated back to the sender. The dispatch and combine phases feature \textit{irregular}, \textit{fine-grained} communication.
    }
    \label{fig:moe-communication-pattern}\vspace{-0.20in}
\end{figure}

MoE models introduce a distinctive communication pattern that differs fundamentally from dense all-reduce or pipeline-parallel patterns found in other models. In a MoE layer, each input token activation is dynamically routed to a small subset of experts based on a learned gating function. This induces a sparse all-to-all pattern: Every GPU holds a subset of experts, so token activations must be dispatched from the originating GPU to the GPUs that host the selected experts and then gathered back to their origin (Figure~\ref{fig:moe-communication-pattern}). The destination GPUs are determined based on the learned gating function during runtime. Every MoE layer's forward pass has two communication phases: \textit{dispatch}, where activations are sent to expert GPUs, followed by a \textit{combine} phase, where expert outputs are returned and merged in the original token order. Expert-parallel communication has several new characteristics:

\paragraph{Fine-grained token-level\footnote{For simplicity, we use token and token activation interchangeably.} transfers.}
In MoE dispatch and combine, each token activation is small, on the order of 7KB (e.g., FP8, hidden size of 7168~\cite{deepseek_v3}), so a naive implementation issues a large number of tiny GPU-GPU transfers. Unlike traditional collective communication (e.g., large, batched all-reduces or all-to-alls) that aggregates data into large bandwidth-efficient messages, MoE communication is naturally fragmented. Traditional approaches must therefore either push many small work-queue entries to the NIC, or first pack activations into contiguous per-expert buffers on the GPU, which consume SM cycles and add packing latency on the critical path.
\paragraph{Irregular communication.}
The number of tokens routed to each expert also varies every iteration, because routing decisions are computed at runtime by the MoE gating network. In contrast, traditional collectives assume a fixed set of participants and largely predictable, symmetric message sizes, which allows for static, precomputed communication schedules. However, MoE communication breaks these assumptions: dynamic routing creates significant load imbalance and introduces substantial overhead from fine-grained, data-dependent transfers~\cite{gale2023megablocks, deepseek-ai_EPLB_2025}. This irregularity makes it hard to precompute efficient communication schedules or reuse static buffer layouts, forcing implementations to dynamically size and route messages every step. This also means load imbalance and incast can pose a challenge, where the destination rank needs to process a burst of arrival tokens. 


\subsection{Expert-parallel communication requires GPU-initiated token-level communication}
\label{ssec:ep_fine_grained_comm}

\begin{figure}[!t]
    \centering
    \includegraphics[width=\linewidth]{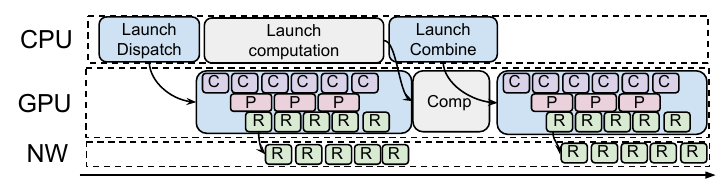}
    \vspace{-0.10in}
    \caption{GPU-initiated token-level communication in DeepEP (High-throughput mode, single batch for illustration). $C$ stands for intra-node data copying via NVLink, $P$ stands for processing, and $R$ stands for RDMA communication. Different phases (computation, communication, and copying) interleave. 
    \vspace{-0.20in}
    }
    \label{fig:deepep_pipeline}
\end{figure}

Given the central role of MoE communication, recent specialized systems—most notably DeepEP~\cite{deepep2025}—have introduced \textit{GPU-initiated token-level communication}, as illustrated in Figure~\ref{fig:deepep_pipeline}. 
    This design involves GPU threads \textit{directly} submitting transfer commands to the NIC, using NVIDIA IBGDA (InfiniBand GPUDirect Async)~\cite{ibgda}. GPU-initiated communication enables fine-grained and pipelined overlap \textit{on a token basis}, where transfer for a single token or a chunk of tokens can overlap with other \textit{phases} of communication, such as data copying between application tensor buffers and RDMA transport buffers, token forwarding from the RDMA domain to the NVLink domain, and necessary computation steps such as the reduce during the combine phase. 
    By breaking communication into these smaller GPU-triggered units (\eg, per-token to 32 tokens), the system can better utilize both network and compute resources and significantly reduce end-to-end communication latency. 

GPU-initiated token-level communication also enables many optimization opportunities, such as message deduplication: if the token activation is routed to experts residing on multiple GPUs \textit{on the same node}, the communication library can only send the token activation once with RDMA, and rely on intra-node forwarding to multiple experts for maximal speed. A second optimization enabled by GPU-initiated token-level communication is hierarchical reduce: an intra-node reduce (weighted sum) is performed on each node for a \textit{chunk of tokens}, the result is sent back to the sender rank for another inter-node reduce: all of which overlaps with background network transfer. Such optimization techniques were previously not feasible, 
and have enabled a significant reduction in the amount of traffic needed to send over the network, and improvement in end-to-end performance.  

\begin{figure}[!t]
    \centering
    \includegraphics[width=0.48\linewidth]{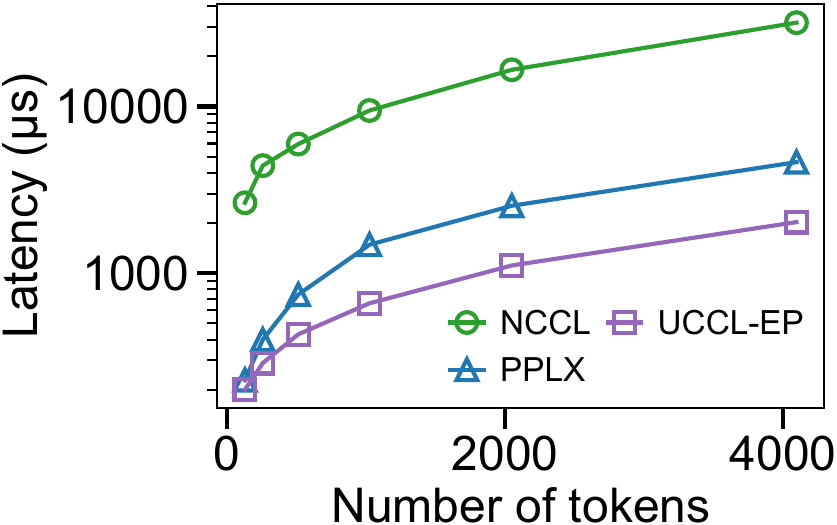}
    \vspace{-0.10in}
    \caption{GPU-initiated token-level communication outperforms coarse-grained bulk transfer (\eg, packing tokens into a contiguous buffer on GPU then CPU initiating a single contiguous transfer) on \pfive (testbed details listed in Table \ref{tab:eval_testbed}). The y-axis is in log scale.}
    \vspace{-0.20in}
    \label{fig:nccl-bad-performance}
\end{figure}

To compare, some inference and training frameworks adopt coarse-grained transfer, such as with NCCL~\cite{nccl} or RCCL~\cite{rccl}, or other general-purpose collective libraries.
They require either the application packing tokens into a contiguous per-destination-rank transfer buffer or transferring small tokens one by one. The former incurs a high overhead of packing the token; the latter suffers from limited transfer throughput with small messages. For example, PPLX~\cite{pplx_garden} adopts on-GPU token packing without fine-grained token deduplication and hierarchical reduce, and it does not scale as the number of tokens increases \amd{(Figure~\ref{fig:nccl-bad-performance})}. 

We summarize existing systems in Table~\ref{tab:existing_solu}. Collective communication libraries such as NCCL and RCCL are designed for regular collective patterns and do not target fine-grained token-level EP. Systems such as DeepEP~\cite{deepep2025} and ROCm-DeepEP~\cite{ROCmDeepEP} support GPU-initiated token-level communication but assume specific GPU-NIC pairings, limiting their portability across heterogeneous platforms.

\subsection{Existing GPU-initiated solutions couple GPUs and NICs, harming portability}

While GPU-initiated token-level communication has been shown to improve performance, unfortunately, it compromises portability by \textit{tightly coupling} GPUs and NICs. The reasons are two-fold:

\paragraph{Lack of hardware interoperability. } The mainstream mechanism of GPU-initiated token-level communication is GPU threads directly posting to the RDMA NICs, or known as IBGDA~\cite{ibgda}. IBGDA requires InfiniBand-capable NICs
that support GPUDirect RDMA so the NIC can directly communicate with GPU memory. It also depends on a compatible software stack to enable GPU-initiated network operations. \amd{Open-source implementations demonstrate that such designs work well on platforms where GPUs and NICs are tightly integrated, for example when NVIDIA~\cite{deepep2025} or AMD~\cite{ROCmDeepEP} GPUs are paired with Mellanox CX-series NICs. However, to the best of our knowledge, no public repository demonstrates that GPU-initiated EP communication achieves comparable performance on other NICs, such as Broadcom Thor2, AWS EFA, or cloud NICs with different transport semantics. As a result, existing GPU-initiated EP systems remain difficult to deploy across heterogeneous GPU and NIC combinations.} 

\begin{table}[t!]
\begin{center}
{\footnotesize
\aboverulesep=0ex
\belowrulesep=0ex
\setlength{\tabcolsep}{1pt}
\begin{tabular}{ccccc}
    \toprule
    &\makecell{Support\\hete GPUs}& \makecell{Support\\hete NICs}&\makecell{Token\\dedup\&reduce}&\makecell{High perf w/\\small tokens}\\
    \toprule
    \makecell{NCCL/RCCL~\cite{nccl, rccl}} &\cmark&\cmark&\xmark&\xmark\\
    \makecell{PPLX~\cite{pplx_garden}} &\xmark&\cmark&\xmark&\cmark\\
    \makecell{DeepEP~\cite{deepep2025}} &\xmark&\xmark&\cmark&\cmark\\
    \makecell{ROCm-DeepEP~\cite{ROCmDeepEP}} &\xmark&\xmark&\cmark&\cmark\\
    \makecell{CPU-asst IBGDA~\cite{ibgda_cpu_assisted}} &\cmark&\xmark&--&--\\
    \hline
    \makecell{\sysname} &\cmark&\cmark&\cmark&\cmark\\
    \toprule
\end{tabular}}
\end{center}
\vspace{-0.1in}
\caption{Comparison of major existing EP communication systems.}
\vspace{-0.2in}
\label{tab:existing_solu}
\end{table}

\paragraph{GPUs impose strict delivery semantics on NICs. } While GPU threads are good for massive parallelism, they are limited in terms of the ability to control and manage the transfer. 
These communication libraries typically expect strict delivery semantics (\eg, ``write-then-atomic'' patterns, applying an atomic only after $x$ number of writes have been delivered). 
This requires the networking layer to respect such delivery semantics, such as providing ordering guarantees. 
Furthermore, GPU threads lack the flexibility to manage communication. For example, DeepEP issues one-sided writes and relies on the receiver to busy poll a flag to detect incoming transfer. 
Congestion (such as incast) can be an issue since the sender is unaware of the message delay, and the current generation of RDMA NICs typically provides subpar congestion control in hardware, requiring a software approach~\cite{meta-training, uccl_collective}. 

Taken together, this means that the GPU-initiated communication introduces strict requirements on the underlying 
networking layer, relying on it to provide reliability, ordering, as well as efficient congestion control. 
Unfortunately, maintaining them not only restricts a narrow selection of networking devices, as well as increasing the necessary fixed cost of enforcing these guarantees in hardware. 

As a result, GPU-initiated token-level communication cannot be supported on cloud NICs (\eg, AWS EFA~\cite{efa} with SRD transport~\cite{aws_srd}) that lack ordering guarantees. As GPU threads have little visibility into transfer delays and network congestion, this also poses significant challenges to practical deployment, as the networking environment needs to be carefully tuned to ensure congestion does not occur~\cite{deepep2025}. 

\paragraph{Importance of portability.}
Portability is essential for cost efficiency and avoiding vendor lock-ins. Prior works have demonstrated significant performance and cost benefits of using heterogeneous hardware~\cite{griggs2024m, jaiswal2025serving, jiang2025hexgen, mei2025helix, li2025taming, zhang2025cauchy, mao2025skyserve}. 
Contemporary large-language-model systems increasingly rely on heterogeneous hardware: different data-centers and cloud providers deploy different NICs with distinct transport protocols (\eg, RC vs. SRD) and capabilities. As such, achieving portability in the communication layer is critical to reducing cost and improving performance, simplifying integration into existing inference or training engines, and making efficient expert-parallel communication usable for a wide range of users. 

\subsection{The challenges of designing a portable and performant expert-parallel communication system}
\label{ssec:challenges-of-portable-ep}

Designing a portable system requires breaking the coupling between GPUs and NICs. \sysname aims to design a portable architecture that enables GPU-initiated token-level communication. We describe the main challenges in the following. 

\paragraph{Heterogeneous GPUs and NICs.}
Both GPUs and NICs are heterogeneous, coming from different vendors with different software ecosystems, \eg, NVIDIA, AMD, AWS EFA, Broadcom, Intel. To enable the GPU to directly initiate communication to the NIC (without involving the CPUs) often requires \textcolor{black}{GPU writes to NIC driver/MMIO interfaces}. This is, at best, error-prone and difficult to do cross-vendor, and more often simply not supported. 

\paragraph{Delivery semantics guarantees. }
Hardware transports differ in whether they guarantee in-order delivery of messages. For example, the ConnectX~\cite{cx7} RC transport offers in-order semantics, while the EFA SRD protocol offers reliable but unordered delivery.  When the GPU kernel assumes ordering guarantees from the networking layer (for instance, issuing operations in a strict sequence without additional synchronization) or requiring a certain group of messages to arrive before a control message is delivered, moving to a transport that delivers out-of-order breaks correctness. A portable communication architecture, therefore, must assume minimal guarantees on the networking layer, or better, allow easy configuration in the software layer to adapt to the heterogeneous networking layer. 

\paragraph{Existing solutions remained specific to each GPU-NIC. }

We observe that existing GPU-initiated communication libraries have remained largely ad hoc and involve one-time solutions on a particular combination of GPUs and NICs. 
\amd{ROCm port of DeepEP~\cite{ROCmDeepEP} is a specialized library for RDMA and GPU integration that works primarily on AMD GPUs with NVIDIA NICs.} CPU-assisted IBGDA~\cite{Langer2024_NVSHMEM3} relies on a single CPU proxy to relay transfer commands from NVIDIA GPUs, suffering from scalability issues, and it can only work on NVIDIA NICs that provide strict ordering guarantees. These solutions require a ported solution for each GPU and NIC vendor pair, inflating the development cost. 
We provide a more exhaustive discussion on the related works in \S\ref{sec:related-works}.

We argue that the expert-parallel communication system should be portable by \textit{design} - and \sysname is based on a set of simple \textit{primitives} that ease portability across various hardware, as well as supporting new future GPUs and NICs with minimal additional overhead. 

\section{Design}

\begin{figure}[!t]
    \centering
    \includegraphics[width=\linewidth]{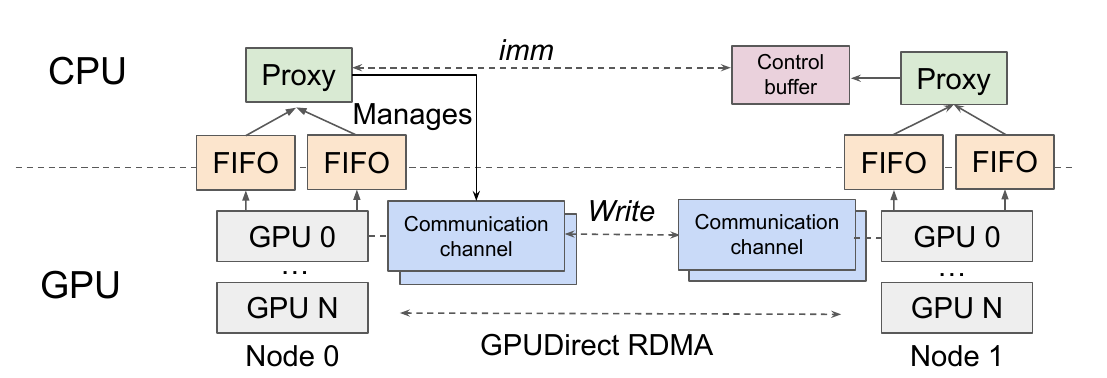}
    \vspace{-0.20in}
    \caption{\sysname architecture. Control buffer temporarily buffers control messages (e.g., atomics) until the conditional check specified by the message is passed, upon which the values carried by these control messages are applied. Multiple communication channels are used for moving the data payloads (\eg, tokens) via GPUDirect RDMA.} 
    \vspace{-0.15in}
    \label{fig:uep-architecture}
\end{figure}

The key observation of \sysname is that the GPU only needs to initiate token-level transfers for maximal performance with fine-grained overlapping with other phases of communication (e.g., data copying and computation), while the responsibility for monitoring and managing those transfers does not need to remain on the GPU. CPU, on the other hand, is portable to any GPUs via CUDA/ROCm and any NICs via the \texttt{libibverbs} library~\cite{libibverbs} maintained by the Linux community and NIC vendors, and CPU is also flexible in terms of enforcing various delivery semantics expected by the GPU. 

In \sysname (Figure~\ref{fig:uep-architecture}), GPUs initiate the \textit{fine-grained} data-movement (\eg, of expert tokens) by delegating the transfer tasks to CPU proxies. In effect, GPU threads issue lightweight control messages over PCIe to the CPU; the CPU proxies then intercept these control messages and issue token-level data communication on behalf of the GPU threads. This separation gives us the best of both worlds: the GPU retains token-level and pipelined data communication necessary for high performance, while the CPU proxy abstracts away heterogeneous NIC semantics with portability across platforms. 

We discuss two high-level primitives proposed in \sysname---an efficient CPU-GPU communication channel (\S\ref{ssec:cpu-gpu-comm-channel}) and a multi-threaded CPU proxy to delegate and manage GPU-initiated communication (\S\ref{ssec:cpu-proxy-threads} and \ref{ssec:express_comm}). 

\subsection{Efficient CPU-GPU communication channel}
\label{ssec:cpu-gpu-comm-channel}

\begin{figure}[!t]
    \centering
    \includegraphics[width=0.7\linewidth]{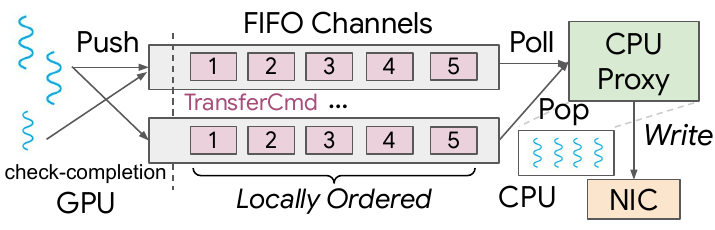}
    \vspace{-0.1in}
    \caption{CPU-GPU FIFO channel. The FIFO channel transmits 128-bit \texttt{TransferCmd}. The GPU threads (left) enqueue commands to the FIFO channel. The CPU proxy threads (right) either read the commands with \texttt{Poll} or dequeue the command with \texttt{Pop}. \sysname employs multiple FIFO channels per GPU, and each CPU proxy uses multiple threads to read from FIFO channels for scalability. }
    \vspace{-0.2in}
    \label{fig:cpu-gpu-communication-channel}
\end{figure}

\sysname employs a lightweight, fixed‐size command descriptor, termed \texttt{TransferCmd}, that the GPU threads enqueue into shared, lock-free FIFO channels with the CPU proxy as consumers (Figure~\ref{fig:cpu-gpu-communication-channel}). The GPU side writes to the tail of the FIFO channel and then proceeds with data packing and forwarding, while the CPU side reads from the head of the FIFO buffer and issues the corresponding RDMA work request to the NIC. 

Each \texttt{TransferCmd} bundles control information necessary for initiating GPU-direct communication, such as destination peer, buffer address, length, and sequence number. This leaves the data payload still residing on the GPU memory. This approach decouples transfer initiation from transfer management.

\paragraph{\sysname FIFO channels.}

Inspired by MSCCL++~\cite{ShahJLRHJMSCZDMY2025}, \sysname uses a 128-bit (16 bytes) descriptor per command to improve the speed of each transfer, as 16 bytes can be written with a single GPU instruction and MMIO doorbell. 
\sysname also caches the tail index value on the GPU, so that when GPU threads look up the tail value, it does not have to cross the PCIe to read it. 
Since MSCCL++ does not target EP and it assumes strict ordering for the NICs, \sysname is significantly different: \sysname uses multi-threaded proxies and multiple parallel FIFO channels to make it scalable for small messages (Figure~\ref{fig:fifo_latency}); 
\sysname further leverages the CPU proxy to enforce delivery semantics among \texttt{TransferCmd}s among a subset of FIFO channels as specified by the GPU threads (\S\ref{ssec:express_comm}) for heterogeneous NICs.

The CPU-GPU channel exposes both CPU-side APIs and GPU-side APIs. \sysname bounds the channel size with a parameter \texttt{kMaxInflight}. A message can be enqueued from the GPU only if there is space in the buffer (the number of messages is equal to or smaller than \texttt{kMaxInflight}) -- otherwise, it will have to wait for the message to be dequeued by the CPU. Therefore, the channel size serves as a way to enforce \textit{control} onto the GPU-initiated communication; this is important for pacing the GPU sender rate. The GPU threads can also, through checking the FIFO channel, know the completion of a prior message (\eg, important for implementing a barrier).

\sysname's CPU-GPU communication channel is \textit{one-sided}:  communication only flows from GPU to CPU
. However, the CPU can rate-limit the GPU by controlling how quickly the messages are consumed from the FIFO buffer. Delaying consumption of the transfer command will make the buffer full, making the GPU thread pause on enqueueing more messages onto the FIFO channel.

The communication channel is split between the host and GPU memory to place each portion of the metadata (e.g., head of the FIFO channel on CPU and tail of the FIFO channel on GPU) where it is most efficiently accessed. This design ensures that the CPU can readily access host-resident state while the GPU accesses device-resident state. 
An alternative design is placing both head and tail on either GPU memory or host memory. However, one side (\eg, CPU or GPU) has to incur repeated access over PCIe when busy polling, incurring multiple PCIe crossings and hence heavy latency cost. 

\paragraph{Memory consistency.} Memory consistency is a key challenge for an efficient CPU-GPU communication channel. If the GPU sees stale data, this can lead to incorrect decisions (\eg, being stalled on a barrier). If the CPU sees stale data, it might read a stale value in the buffer (leading to incorrectness). \sysname ensures that when the GPU accesses the channel, it will bypass its hardware caches. CPU-side also ensures that the writes are flushed to the host memory rather than buffered in its L2 cache. Alternatively, the cache-coherent C-C interconnect (e.g., in GH200~\cite{nvidia_gh200_grace_hopper_superchip}) can mitigate this issue (Figure~\ref{fig:gh200_perf} in evaluation). 

\paragraph{Impact of GPU-side contention.} Due to the large number of GPU threads, it is common that multiple threads of the same GPU SM try to push commands into the same FIFO channel. To reduce contention, \sysname uses multiple FIFO channels and carefully routes each GPU thread to push to the relevant FIFO channel. Cross-FIFO channels ordering is not guaranteed, though commands enqueued within the same FIFO channels are guaranteed to be ordered. \sysname ensures that the sequence of messages that the GPU kernel requires ordering from the GPU side is mapped to the same FIFO channel.

\paragraph{Channel APIs.}

We next discuss APIs exposed at both the CPU-side and GPU-side to enable GPU-initiated token-level expert-parallel communication.

\textit{\underline{CPU-side: }}
\begin{itemize}
    \item \textbf{Poll}: CPU reads the \texttt{TransferCmd} from the channel, without dequeueing the command from the channel. 
    \item \textbf{Pop}: CPU removes the \texttt{TransferCmd} from the channel, indicating that the \texttt{TransferCmd} has been consumed.
\end{itemize}

\textit{\underline{GPU-side: }}
\begin{itemize}
    \item \textbf{Push}: GPU thread enqueues a \texttt{TransferCmd} into the FIFO channel and gets a \texttt{Idx} for the command. 
    \item \textbf{Check-completion(Idx)}: The GPU thread checks whether the \texttt{TransferCmd} with \texttt{Idx} has been popped from the CPU side. 
\end{itemize}

\sysname decouples between initiating the transfer and finishing the transfer. The CPU first becomes aware of the transfer message when calling \texttt{Poll}, reading the message from the FIFO queue. This allows the CPU proxy thread to initiate a particular operation. For example, the CPU proxy thread can issue an RDMA write after polling a \texttt{Write} \texttt{TransferCmd} (described later). \sysname gives the CPU proxy the flexibility to either notify the GPU when the operation has completed, such as receiving the completion queue entry for that write, or immediately notify the GPU when the operation initiates. The choice to either notify the GPU immediately when the operation initiates or to delay notifying the GPU after completion depends on the message types. For example, if the networking transport protocol guarantees reliability (where messages cannot be lost), then the \texttt{Write} \texttt{TransferCmd} can be \texttt{Pop}-ed immediately conditioned on some constraint (such as the number of current inflight messages is under a limit, \texttt{kMaxInflight}). However, for other message types, such as establishing a synchronous barrier (Barrier \texttt{TransferCmd}), the \texttt{TransferCmd} can only be removed from the FIFO queue after a synchronous barrier has been successfully established. 

The FIFO queue exposes an API for the GPU thread to check completion. This is important for GPU-side operations. For example, the GPU-side might need to wait for prior dispatch or combine operations to finish to proceed; this logic often requires the communication layer to initiate a barrier across a subset of participating ranks. Therefore, the GPU-side needs to wait for the barrier operation to finish by checking completion on the Barrier \texttt{TransferCmd}. 

\paragraph{Types of TransferCmd.}
We added four types of \texttt{TransferCmd}s to support GPU-initiated token-level communication. These types are not meant to be an exhaustive list for expert-parallel communication; we use DeepEP as an example to highlight the general approach of using FIFO channels and CPU proxy threads to bridge the delivery semantics between GPUs and the underlying networking layer.

\textbf{Write}: GPU thread delegates a write request to the CPU proxy thread to execute. We pass in the \texttt{addr offset} on the source rank, \texttt{addr offset} on the destination rank, the number of bytes to transmit, and the destination rank. Only offsets are needed, rather than a global address, as \sysname CPU proxies exchange each other's base address during initiation (\S\ref{ssec:cpu-proxy-threads}). 
The write command can optionally piggyback an atomic message to signal completion. This can be useful when GPU threads both want to deliver a data payload to the destination memory address, as well as incrementing a related counter (\eg, the number of the payload received) at the receiver. 

\textbf{Atomics}: GPU thread delegates a standalone atomic operation to the CPU proxy thread to execute. The request contains destination offset, atomic value, and destination rank. This operation should execute atomically with concurrent atomics from both senders on different nodes as well as on the same node. Different networking vendors have varying support for atomics; we describe one example in \S\ref{ssec:efa-case-study}  on EFA. Atomics are used to support a variety of higher-level semantics, such as acting as a remote doorbell to ensure previous writes are completed and to update the head and tail of the communication channel, which is implemented as a ring buffer (\S\ref{ssec:express_comm}).  

\textbf{Drain}: Delegates a CPU proxy thread to drain the RDMA completion queue and ensure that all outstanding RDMA operations have completed. This is critical to ensure that the GPU threads can safely proceed without having their local states overwritten by unfinished requests from previous iterations of dispatch and combine. \sysname optionally allows the GPU threads to pass a parameter in the Drain command to drain in-flight messages up to a certain message index \texttt{Idx}. 

\textbf{Barrier}: Barrier message delegates the CPU proxy thread to establish a synchronization barrier: the barrier message can support an all-peer barrier (where all participating peers need to enter the barrier before exit), or a same-rail barrier that, under a rail-optimized topology, synchronizes only the peers mapped to the same rail (same GPU index). For the former, the \sysname Proxy thread creates an intra-node shared memory to enforce a hierarchical barrier---synchronization among local peers first, then across nodes. A per-node leader is established for intra-node barrier, and a cross-node leader (\eg, typically the first node) is established then for inter-node synchronization.  For the latter, \sysname relies on the RDMA immediate data to signal the leader rank that a barrier request is received, then the leader rank replies with the immediate data to signal the other ranks that a barrier has been successfully established.

Both Barrier and Drain are blocking on the GPU side, meaning that the GPU thread will need to check completions with the correct \texttt{TransferCmd Idx} before it can proceed with the \texttt{Check-completion(Idx)} API. While more message types can be supported, \sysname supports the above four as they are adequate to cover the full functionality of DeepEP. 

\subsection{Flexible CPU proxy to delegate communication}
\label{ssec:cpu-proxy-threads}

\sysname keeps one CPU proxy per GPU, and each CPU proxy has multiple threads. Different CPU threads do not share state and require no synchronization. While the use of CPU threads is common in collective transfers, such as in NCCL~\cite{nccl} and CPU-assisted IBGDA~\cite{ibgda_cpu_assisted}, \sysname differs fundamentally from these approaches in three key aspects. First, the communication pattern is GPU-initiated and \textit{token-level}, posing challenges for CPU proxy threads to quickly handle and process messages from GPUs. Second, as the message size is smaller, the number of messages needed to saturate network bandwidth is substantially higher; \sysname uses more CPU threads for scalability. Third, \sysname's CPU proxy is tasked with bridging the delivery semantics expected by the GPU (\S\ref{ssec:challenges-of-portable-ep}) and provided by the networking layer, and efficiently expressing these requirements without hurting performance is a non-trivial problem. 

\paragraph{Establishing connection.} 
With multiple threads in a CPU proxy, \sysname bounds the number of connections by only allowing the $i$-th thread of a proxy to establish connections with the $i$-th threads of remote proxies. 
Communication is bidirectional: each thread is in charge of both polling the sender's completion queue for any successful outgoing write or atomic message, and polling the receiver's completion queue to handle any remote incoming requests. 

\paragraph{Symmetric memory.} The CPU proxy thread maintains the \textit{illusion} of symmetric memory. Symmetric memory has been noted to be useful by prior works~\cite{deepep2025, pytorch_symmetric_memory, nvidia_nvshmem}. 
This enables the GPU thread to only notify CPU threads of the \textit{offsets} of the transfer. CPU threads then handle the address translation, bounds checking before issuing RDMA writes, and atomic messages. Each CPU proxy registers a memory region during initialization, and exchanges the base address during handshake. Using offsets in symmetric memory reduces the number of bits needed in control messages, compared to passing full addresses. This also eliminates the need to use vendor-specific shared memory libraries, such as NVSHMEM~\cite{nvidia_nvshmem} and rocSHMEM~\cite{rocshmem}, improving portability. 

\paragraph{Addressing delivery semantics.}

Heterogeneous NICs exhibit different delivery semantics: for example, some NICs may deliver RDMA writes out of order. 
To handle this, the CPU proxies embed a sequence number in each RDMA write via immediate data; the receiver uses the sequence number to reorder or delay processing the control messages until all prior writes have arrived. 
Immediate data is a 32-bit piece of data that RDMA send/write operations can piggyback in the packet header and deliver to remote CPUs over the network. 
By having CPU interpret this immediate data and enforce guarantees, the system remains correct even when the transport does not guarantee strict delivery guarantees. For comparison, in IBGDA, all transfers \textit{completely} bypass the CPU. We delve into the details of this design next. 

\subsection{Expressing GPU communication requirements with CPU proxy}
\label{ssec:express_comm}

Similar to DeepEP~\cite{deepep2025}, \sysname presents two modes: low latency (LL) mode and high-throughput (HT) mode. Low-latency mode is used for the decode phase with a smaller batch size, and high-throughput mode is used for the prefill phase or training, where there are more tokens in the batch. LL mode immediately sends the token activation initiated by the GPU via the CPU proxy, requiring no synchronization between transfers. HT mode implements message deduplication and intra-node forwarding. High-throughput kernel employs multiple communication channels (Figure~\ref{fig:uep-architecture}), a set of ring buffers per GPU that temporarily buffer tokens to send in the granularity of chunks (a configurable parameter, typically 32 tokens). 

In the remainder of the subsection, we use \sysname's low-latency (LL) kernel and high-throughput (HT) kernel as two illustrative examples of \sysname's proxy threads enforcing delivery semantics required by GPU kernels. Each messages (both write and atomic) is tagged with a 32-bit immediate value. Note that RDMA writes are still immediately applied, though the receiver CPU proxy retrieves an immediate data through polling the completion queue. Atomic is implemented via the immediate data and is not immediately applied: CPU proxy thread extracts the offset and value for the atomic itself from the 32-bit immediate data, buffers them in the control buffer (Figure~\ref{fig:uep-architecture}) if needed, and selectively applies them based on the delivery semantics expected of these atomics. The delivery semantics are dictated by the higher-level protocol these atomics implement (e.g., a remote-completion doorbell).

\paragraph{Low-latency kernel requires partial completion fence.}
Atomic messages are used to signal token delivery (\eg, $X$ number of writes have completed), requiring that the receiver wait until the required number of writes for a specific expert has finished. This requires \textit{completion fence semantics}: if atomic arrives before $X$ tokens, it should not be applied. However, it does not matter if these $X$ tokens are delivered in-order. This guarantee is \textit{partial}: completions of writes to other experts do not affect updating the number of delivered tokens of this expert. 

\textit{\underline{Solution:}} \sysname lets the CPU proxy temporarily buffer the atomic message in the control buffer  until the required number of tokens for a destination expert has been received. To achieve that, \sysname packs the destination offset and the expert index into the 32-bit immediate data, and the receiving CPU thread will parse the immediate data, extract the expert index as well as the source rank of the connection. Each subsequent atomic message will go through a lightweight conditional check (has $X$ number of writes to the specified expert being received?): \sysname only applies the atomic update when the conditional check has passed. 

\paragraph{High-throughput kernel requires partial ordering.}

The high-throughput kernel employs multiple ring buffers as communication channels per GPU to temporarily buffer tokens to send. Each token is written to a slot in the destination ring buffer, and by carefully controlling the head and tail values of the ring buffer, \sysname ensures that no token will overwrite other yet-to-be-read tokens on the same slot of the ring buffer. 
A write message is typically followed by an atomic operation to increment either the head or tail values. If the write and atomic become reordered with other writes and atomics, this can lead to the receiver reading stale data from the communication channel, or the sender overwriting data written by a prior transfer. 

Similar to LL mode, enforcing ordering only needs to be partial. 
The ordering guarantees only need to be done for each communication channel, rather than globally across all messages, which is expensive if not done in hardware.  

\textit{\underline{Solution:}} \sysname ensures that per-channel communication is locally ordered. It does this by enqueueing the messages from the same communication channel to the same FIFO queue. 
Similar to the low low-latency kernel, the receiver CPU proxy thread will extract the sequence number from the immediate data; if the received message arrives out-of-order, it will temporarily buffer these atomic messages (that are out-of-order with writes) in the control buffer. Only after the prior writes and atomics have been applied (e.g., message $1-5$ from communication channel $i$ has been applied), the receiver thread will sequentially apply the next buffered atomic message (e.g., atomic with index $6$ from communication channel $i$). 

\begin{figure}[!t]
    \centering
    \includegraphics[width=0.48\linewidth]{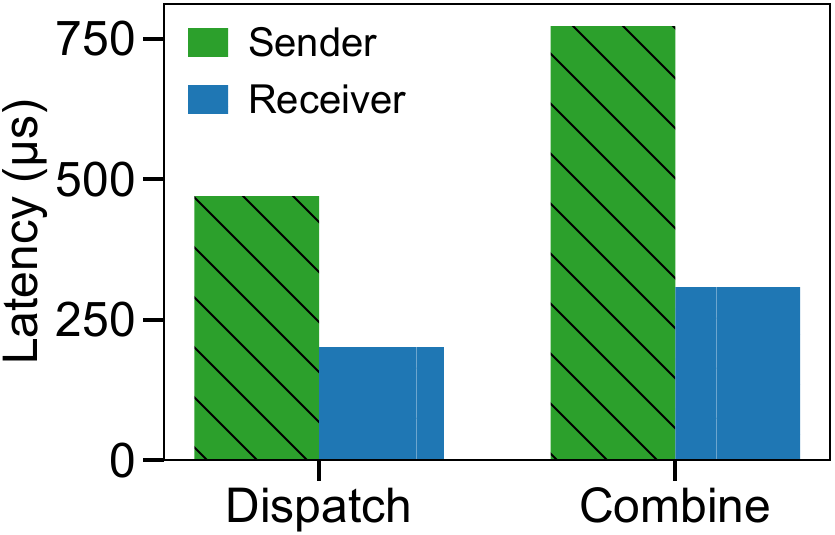}
    \vspace{-0.1in}
    \caption{Sender-side vs. receiver-side on enforcing delivery semantics (on testbed \pfive detailed in Table~\ref{tab:eval_testbed}). \textit{Receiver-side} gives better performance compared to sender-side.}
    \vspace{-0.2in}
    \label{fig:sender-vs-receiver-barrier}
\end{figure}

\paragraph{Where to enforce the delivery semantics.} An alternative approach is for the \textit{sender}, as opposed to the receiver, to delay sending the atomic message only after it has received the completion queue entry for the writes sent for that token.  This approach is typically used by the NIC hardware to guarantee strict ordering when adaptive routing is used in the network and packet reordering happens~\cite{c4_alibaba}. It has the advantage of saving hardware SRAM resources without tracking per-packet states.
However, compared to tracking at the receiver, this approach makes the sender wait for one extra RTT for the atomic sent. We have observed suboptimal performance (Figure~\ref{fig:sender-vs-receiver-barrier}), as the latency penalty of waiting for token completion accumulates.

In summary, the networking layer typically provides either a stronger or weaker guarantee in hardware than what is typically required by the application, \eg, ensuring that \textit{all} messages are strictly ordered, or guaranteeing none at all. From discussing with industry practitioners, installing various guarantees comes at a hardware cost trade-off: supporting strong guarantees at the NIC hardware typically means that the hardware NICs are more expensive to make. In comparison, CPU is flexible in customizing and enforcing various delivery semantics. 
\section{Implementation}

\begin{table*}[t]
\begin{center}
{\footnotesize
\begin{tabular}{cccccccc}
    \toprule
    \makecell{Name}
    &\makecell{\# servers}
    &\makecell{Network}
    &\makecell{GPU}
    &\makecell{HBM, \# SMs, NVLink\amd{/xGMI}}
    &\makecell{NIC}
    &\makecell{CPU}
    &\makecell{Cloud}
    \\\toprule
    \pfive & 4 & Ethernet & NVIDIA H200$\times$8 & 141 GB, 132, 900 GB/s & AWS EFAv3 200G$\times$16 & 192 cores & AWS \\
    \psix & 4 & Ethernet & NVIDIA B200$\times$8 & 192 GB, 160, 1800 GB/s & AWS EFAv4 400G$\times$8 & 192 cores & AWS \\
    \nebius & 4 & InfiniBand & NVIDIA H100$\times$8 & 80 GB, 132, 900 GB/s & NVIDIA ConnectX-7 400G$\times$8 & 128 cores & Nebius \\
    \lambdalab & 2 & InfiniBand & NVIDIA GH200$\times$1 & 96 GB, 132, -- & NVIDIA ConnectX-7 200G$\times$1 & 72 cores & Lambda \\
    \amdib & 4-16 & \amd{Ethernet} & AMD MI300X$\times$8 & 192 GB, 304, 896 GB/s & NVIDIA ConnectX-7 400G$\times$8 & 128 cores & OCI \\
    \amdbrc & 4 & Ethernet & AMD MI300X$\times$8 & 192 GB, 304, 896 GB/s & Broadcom Thor‑2 400G$\times$8 & 128 cores & Vultr \\
    \toprule
\end{tabular}
}
\end{center}
\vspace{-0.20in}
\caption{Evaluation testbeds. All testbeds are rented from public cloud providers. 
}
\vspace{-0.15in}
\label{tab:eval_testbed}
\end{table*}

We implement \sysname by extending DeepEP with 20.8K lines of C++ (including 2.4K lines of CUDA/ROCm C++) and 1K lines of Python, while remaining API-compatible with DeepEP. 
\sysname significantly extends DeepEP in two ways: i) supporting heterogeneous GPUs and NICs, including NVIDIA and AMD GPUs, and NVIDIA, AWS EFA, and Broadcom NICs (other NIC vendors should be naturally supported via the portable \texttt{libibvers}); and ii) supporting token-level and customizable communication requirements across NICs and EP modes (\ie, LL or HT), as described in \S\ref{ssec:express_comm}. 
\sysname is architecturally portable: porting to AMD GPUs and AWS EFA NICs is done with \textit{only} $3$ person-months, requiring relatively less effort compared to supporting such communication across every NIC and GPU pair, which typically requires dedicated teams. 

\paragraph{Removing GPU-vendor-specific software stack.} 
Existing GPU-driven communication stacks often rely on NVSHMEM for device-side synchronization, memory ordering, and GPU-initiated one-sided operations. However, such a software stack assumes a specific GPU vendor and depends on hardware features (\eg, RC ordering, BAR1 mappings) that may not exist on AMD GPUs or non-IB NICs. \sysname instead manages symmetric memory with CPU proxy threads, and expresses NVSHMEM-related GPU-initiated communication APIs via CPU proxy.

\paragraph{Queue Pair (QP) load balancing.} Each proxy thread for a particular GPU is in charge of managing a set of NICs in the same NUMA group as the GPU (\eg, there are 2$\times$200G EFA NICs per H200 GPU on AWS). In low-latency (LL) mode, each thread creates an RDMA queue pair for a given destination rank; in high-throughput mode, each thread creates multiple QPs (corresponding to the number of FIFO queues) between pairs of ranks (in the same rail). For example, the number of QPs per node for EP=32 is $8$ FIFO queues $\times\ 8$ GPUs (on the same rail) $\times\ 4$ nodes $= 256$. To compare, each EFA NIC supports 256 QPs. Depending on communication requirements, the GPU might require a set of messages to be sent out from a single QP, or it does not impose any requirements on which QPs to send, where the CPU thread round-robins among the QPs it manages.

\paragraph{Aggregating NICs of different bandwidths.} Beyond QP load balancing, \sysname aggregates bandwidth across multiple NICs per GPU: \eg, one ConnectX-7 NIC may deliver 400 Gbps, but achieving similar bandwidth with EFA NICs may require aggregating multiple 200 Gbps NICs per GPU. \sysname relies on CPU threads to load balance across different NICs. We omit the details for brevity. 

\subsection{Supporting EFA}
\label{ssec:efa-case-study}

\paragraph{Emulating atomics with CPU proxy threads.}
EFA NICs currently do not provide hardware RDMA atomics (\eg, global counters). \sysname implements software‐based atomics: the sender issues a payload write followed by a small RDMA write carrying an immediate value encoding the new counter or flag; the CPU proxy or receiver thread updates local completion counters allocated on the host memory (\eg, via \texttt{cudaMallocHost()}) upon detection of the immediate data. \sysname carefully ensures that the GPU observes the host-allocated counter memory and uses it for control decisions. This ensures the correctness of completion notification without relying on vendor-specific atomic support, which improves portability and reduces the complexity of the GPU kernels.

\subsection{Quick adaption on AMD}
\label{ssec:amd-case-study}

\sysname generalizes the expert-parallel communication kernels so that they no longer assume NVIDIA-style warps, NVIDIA-specific PTX intrinsics, and hardware engine. In particular, \sysname does the following changes: 
\begin{itemize}[noitemsep, topsep=0pt, leftmargin=*]
    \item Migrating CUDA-specific PTX intrinsics to use ROCm alternatives, including atomics, memory fences, and timers.
    \item Migrating CUDA warp-based programming to support AMD wavefront, including switching \texttt{WARP\_SIZE} from 32 to 64, using AMD wavefront-level synchronizations. 
    \item Migrating NVIDIA TMA-based data copy~\cite{tma_engine} to support AMD CU-based (\ie, compute units, like NVIDIA SMs). 
    \item \sysname uses wavefront (\ie, warp in NVIDIA's term) specialization for AMD. For the DeepEP HT kernel, \sysname merges its \texttt{coordinator}-role wavefronts into \texttt{receiver}-role wavefronts~\cite{deepep_ht}, as AMD GPUs usually support fewer wavefronts than NVIDIA warps but more threads per wavefront. 
\end{itemize}

Note that to support AMD GPUs with heterogeneous NICs, only GPU kernels need to be ported, rather than the CPU-side code to operate on heterogeneous NICs; \sysname's approach allows us to immediately run on AMD platforms with heterogeneous NICs after AMD GPU-side changes, without having to do independent development between AMD GPU and each individual NIC. This shows that \sysname enables $O(m)$ effort, instead of IBGDA's $O(m\times n)$, to support GPU-initiated token-level communication for expert parallelism. 

\section{Evaluation}

Our evaluation aims to answer the following questions: 

\begin{itemize}[noitemsep, topsep=0pt, leftmargin=*]
    \item How does \sysname's performance compare to baselines on heterogeneous devices? (\S\ref{sssec:nvidia_gpu} and \S\ref{sssec:amd_gpu})
    \item How does \sysname's performance compare to the original DeepEP on NVIDIA GPUs and NICs? (\S\ref{sssec:nvidia_gpu}) 
    \item How does \sysname improve MoE model training and serving? (\S\ref{ssec:eval-inference} and \S\ref{ssec:eval-training})
    \item How do different design choices (e.g., number of CPU threads) impact \sysname performance? (\S\ref{ssec:drill_down})
\end{itemize}

\subsection{Methodology}

\paragraph{Experimental setups.} A list of testbeds can be found in Table~\ref{tab:eval_testbed}. We used testbeds from a variety of GPU vendors (NVIDIA and AMD) and NIC vendors (AWS, NVIDIA, Broadcom) to show that \sysname is portable across platforms and benchmark \sysname's performance. 

\paragraph{Baselines.} 
\begin{itemize}[noitemsep, topsep=0pt, leftmargin=*]
    \item \textbf{NCCL~\cite{nccl} / RCCL~\cite{rccl}.}
    We use collective communication libraries as a baseline for the EP communication stack on NVIDIA (NCCL) and AMD (RCCL) GPUs.
    \item \textbf{DeepEP~\cite{deepep2025}.}
    DeepEP is a state-of-the-art, GPU-initiated RDMA communication system for expert-parallel MoE that serves as the performance upper bound on NVIDIA hardware.
    \item \textbf{Perplexity Kernels (PPLX)~\cite{pplx_garden}.}
    We evaluate against Perplexity's custom MoE communication kernels (denoted PPLX), which are highly optimized for low-latency decode. We used its latest version~\cite{pplx_garden} rather than its old version~\cite{pplx-kernels} as the new version has better performance. 

    \item \textbf{CPU-assisted IBGDA~\cite{ibgda_cpu_assisted}.}
    We emulate a CPU-assisted IBGDA design using a single \sysname proxy thread, similar to the existing CPU-assisted IBGDA approach. 
    \item \textbf{Theoretical Best}: Theoretical results for HT mode derived from available RDMA bandwidth. 
\end{itemize}

To ensure a fair comparison, we use the same amount of GPU resources (e.g., the same number of SMs per GPU as DeepEP, the same number of CUs on AMD). 
\sysname uses 4 CPU threads per GPU for communication.
 
\subsection{Microbenchmark} 
\label{ssec:comparison_hete}

In this subsection, we compare the dispatch and combine latency on NVIDIA GPUs (\S\ref{sssec:nvidia_gpu}) as well as AMD GPUs (\S\ref{sssec:amd_gpu}). Under each GPU vendor, we compare different NIC setups, including AWS EFA, CX7 with IB, and Broadcom Thor-2 NICs. \sysname is able to run across all testbeds. In each testbed, we compare \sysname against the available baselines. 

\subsubsection{On NVIDIA GPUs}
\label{sssec:nvidia_gpu}

\begin{figure}[!t]
\centering
    \begin{subfigure}[t]{0.48\linewidth}
        \centering
        \includegraphics[width=\linewidth]{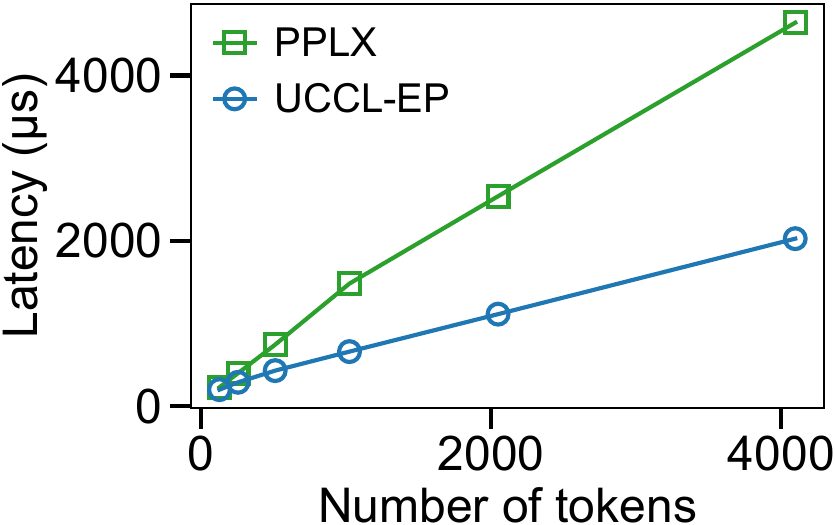}
        \caption{Dispatch.}
        \label{fig:dispatch-latency-across-token}
    \end{subfigure}
    \hfill
    \begin{subfigure}[t]{0.48\linewidth}
        \centering
        \includegraphics[width=\linewidth]{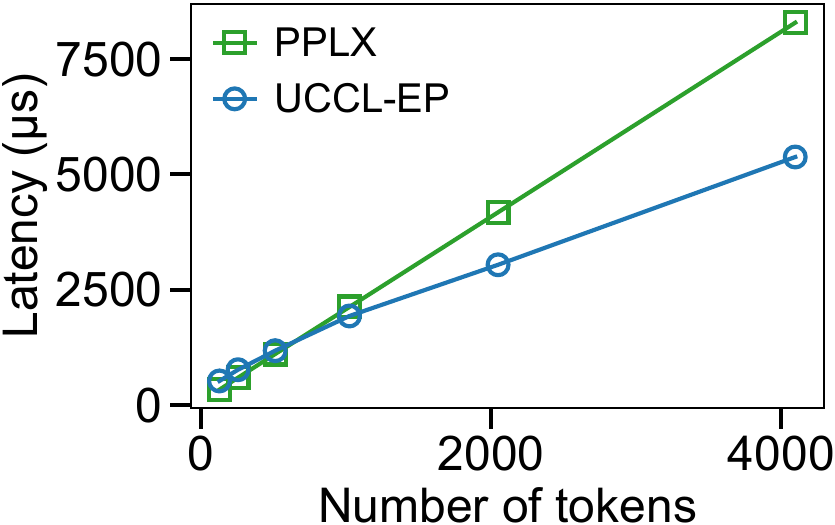}
        \caption{Combine.}
        \label{fig:combine-latency-across-token}
    \end{subfigure}
\vspace{-0.1in}
\caption{EP32 comparison when varying numbers of tokens on \pfive. \sysname uses the minimum of HT and LL latency, while PPLX only has one mode. 
}
\label{fig:eval_overall_p5en}
\vspace{-0.1in}
\end{figure}

\begin{figure}[!t]
\centering
    \begin{subfigure}[t]{0.48\linewidth}
        \centering
        \includegraphics[width=\linewidth]{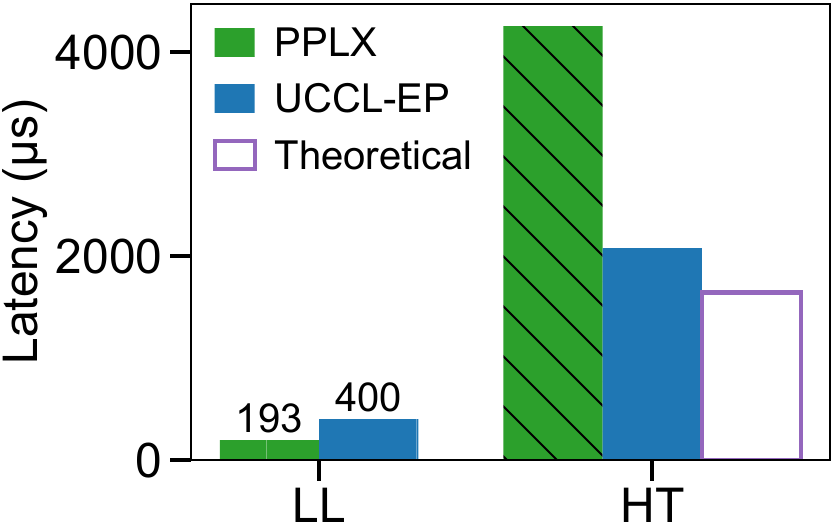}
        \caption{Dispatch.}
        \label{fig:dispatch-latency-p6}
    \end{subfigure}
    \hfill
    \begin{subfigure}[t]{0.48\linewidth}
        \centering
        \includegraphics[width=\linewidth]{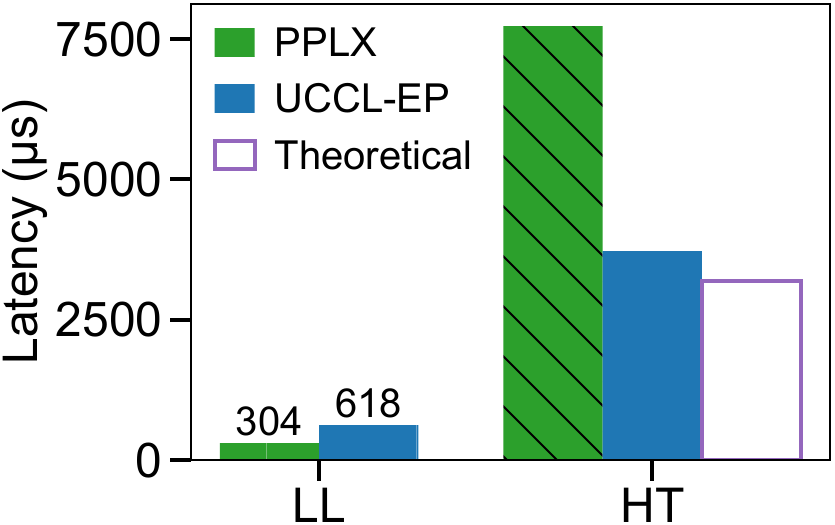}
        \caption{Combine.}
        \label{fig:combine-latency-p6}
    \end{subfigure}
\vspace{-0.1in}
\caption{EP32 comparison on \psix.} 
\label{fig:eval_overall_p6}
\vspace{-0.2in}
\end{figure}

\paragraph{Comparison on AWS EFA.}
Figure~\ref{fig:eval_overall_p5en} and Figure~\ref{fig:eval_overall_p6} show EP32 dispatch and combine latency on NV\_EFAv3 (on AWS \texttt{p5en} instances) as we vary the number of tokens. At small batches (128 tokens), PPLX achieves lower latency than \sysname. This is because \sysname, extending DeepEP, issues messages at 7KB token granularity (current EFA firmware is not able to process small tokens at high rate\footnote{AWS has confirmed that they are working on a firmware fix for this issue.}, Figure~\ref{fig:fifo_latency}), whereas PPLX packs tokens into a larger message. However, as we increase the number of tokens, \sysname quickly overtakes PPLX and the gap widens: for medium and large batches, \sysname consistently delivers substantially lower dispatch ($2.3\times$) and combine latency (1.1-1.5$\times$), demonstrating better scalability as token counts increase. 
 
\begin{figure}[!t]
\centering
    \begin{subfigure}[t]{0.48\linewidth}
        \centering
        \includegraphics[width=\linewidth]{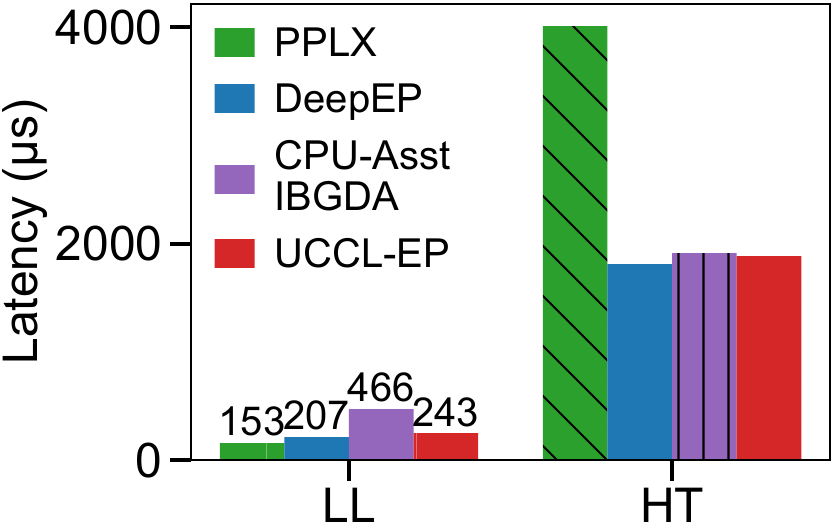}
        \caption{Dispatch.}
        \label{fig:dispatch-latency-nebius}
    \end{subfigure}
    \hfill
    \begin{subfigure}[t]{0.48\linewidth}
        \centering
        \includegraphics[width=\linewidth]{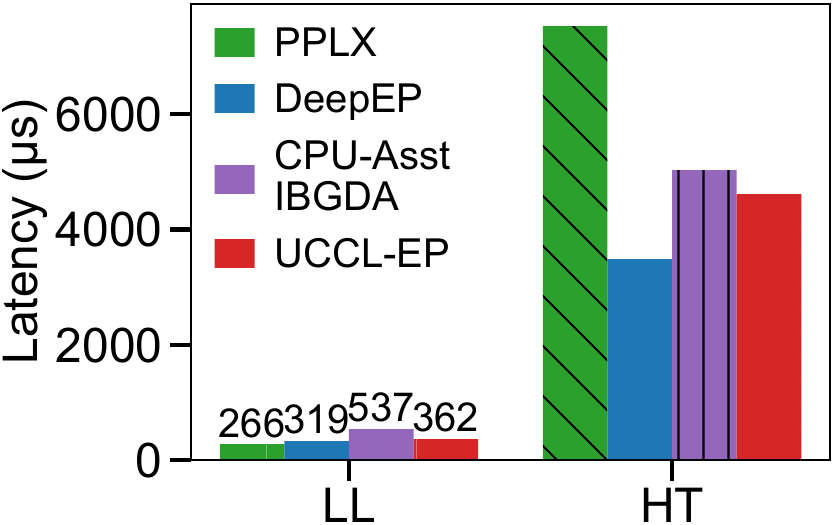}
        \caption{Combine.}
        \label{fig:combine-latency-nebius}
    \end{subfigure}
\vspace{-0.1in}
\caption{EP32 comparison on \nebius. 
}
\label{fig:eval_overall_nebius}
\vspace{-0.1in}
\end{figure}

\paragraph{Comparison on CX7 with IB.}
Figure~\ref{fig:eval_overall_nebius} compares EP32 dispatch and combine latency on CX7 with InfiniBand for both low-latency (LL) and high-throughput (HT) modes. In LL mode, \sysname incurs slightly higher latency than DeepEP and PPLX due to the overhead of its CPU proxy on small messages. However, in HT mode, \sysname achieves latency comparable to the original DeepEP (within $5\%$ for dispatch) while outperforming PPLX for both dispatch ($2.1 \times$) and combine ($1.6 \times $), showing that \sysname preserves DeepEP-level performance on throughput-oriented workloads.

\begin{figure}[!t]
    \centering
    \includegraphics[width=0.48\linewidth]{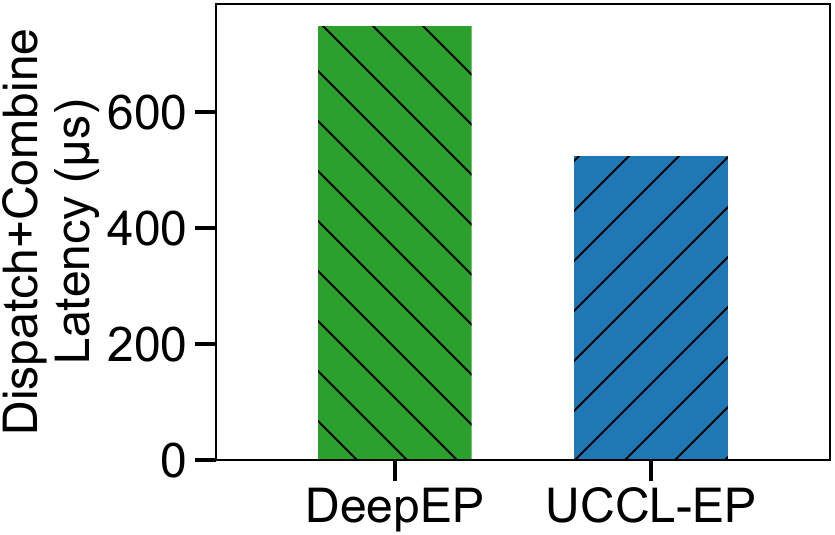}
    \vspace{-0.1in}
    \caption{EP2 LL comparison on two GH200 nodes. While EP2 is a less practical setting, it shows the current trend of unified memory between GPUs and CPUs (e.g. with NVLink-C2C) enables us to obtain better performance by enabling fast CPU-GPU communication. 
    }
    \vspace{-0.05in}
    \label{fig:gh200_perf}
\end{figure}

\paragraph{Comparison on GH200 with NVLink-C2C.}
Figure~\ref{fig:gh200_perf} reports latency on a single-GPU GH200 node with NVLink-C2C between the CPU and GPU. The HT (high-throughput) mode of DeepEP requires an 8-GPU NVLink/NVSwitch topology and therefore cannot run on this platform, so we only compare LL mode. In this setting, \sysname achieves lower transfer latency 
than the original DeepEP, demonstrating that \sysname operates efficiently over a cache-coherent C2C CPU–GPU interconnect; we hypothesize that this benefit comes from removing the NVSHMEM dependency and its associated software overheads. 

\subsubsection{On AMD GPUs}
\label{sssec:amd_gpu}

\begin{figure}[!t]
    \centering
    \begin{subfigure}[t]{0.48\linewidth}
        \centering
        \includegraphics[width=\linewidth]{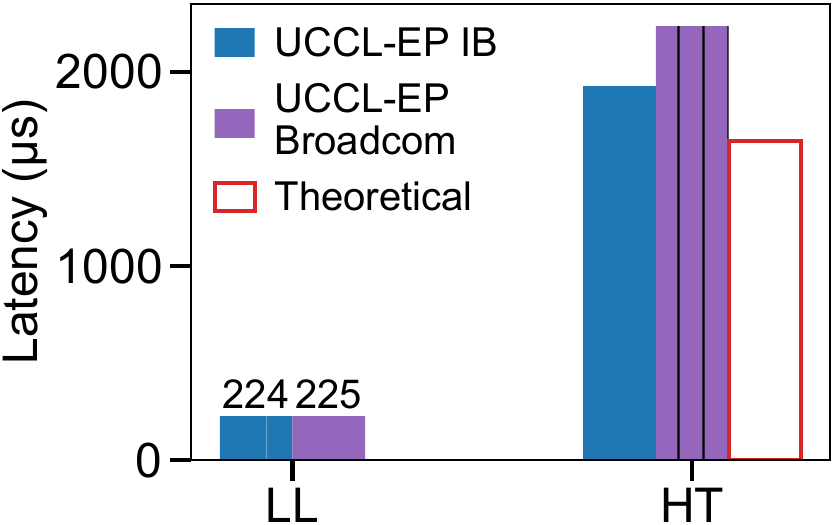}
        \caption{Dispatch.}
        \label{fig:dispatch-latency-amd}
    \end{subfigure}
    \hfill
    \begin{subfigure}[t]{0.48\linewidth}
        \centering
        \includegraphics[width=\linewidth]{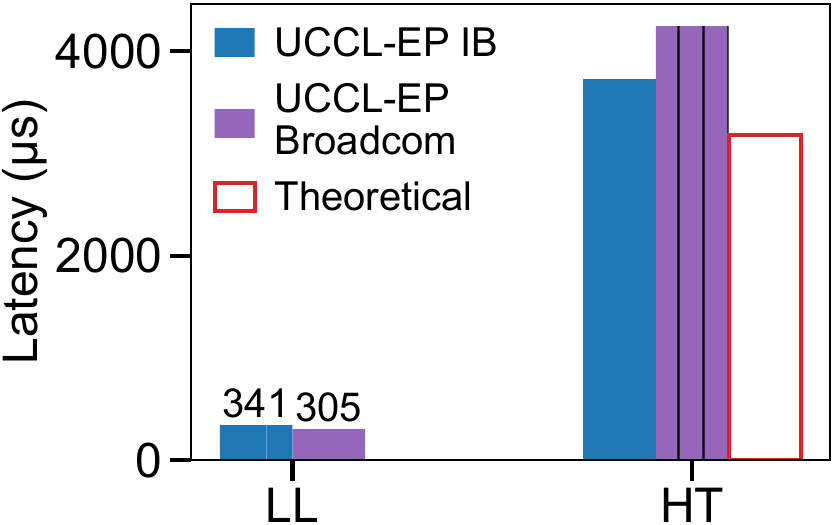}
        \caption{Combine.}
        \label{fig:combine-latency-amd}
    \end{subfigure}
    \vspace{-0.1in}
    \caption{EP32 comparison on \amdib and \amdbrc.}
    \vspace{-0.15in}
    \label{fig:amd_perf}
\end{figure}

Figure~\ref{fig:amd_perf} shows EP32 dispatch and combine latency on AMD GPUs when using Broadcom NICs (\sysname Broadcom) and NVIDIA InfiniBand NICs (\sysname IB). 
The results demonstrate that \sysname runs efficiently across heterogeneous NICs, achieving similar performance on Broadcom and IB in both LL and HT modes. 

\subsection{Application performance}

\begin{figure}[!t]
    \centering
    \begin{subfigure}[t]{0.48\linewidth}
        \centering
        \includegraphics[width=\linewidth]{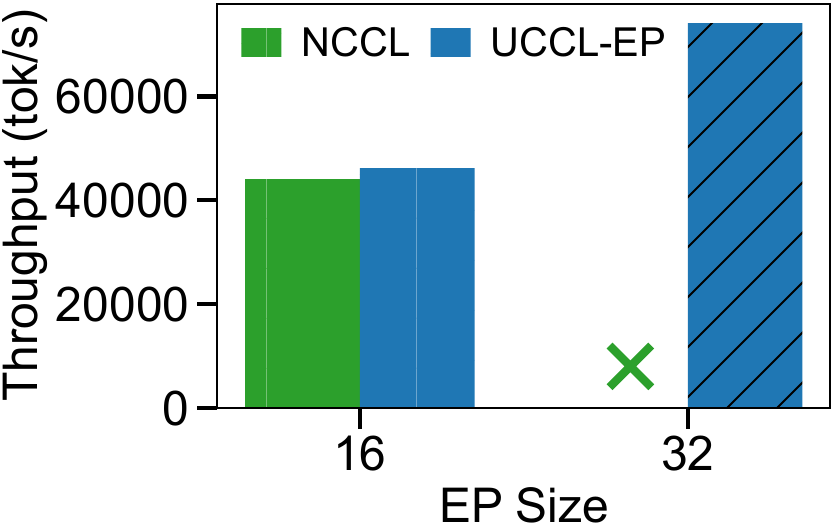}
        \caption{DeepSeek R1 (671B)}
        \label{fig:sglang-r1}
    \end{subfigure}
    \hfill
    \begin{subfigure}[t]{0.48\linewidth}
        \centering
        \includegraphics[width=\linewidth]{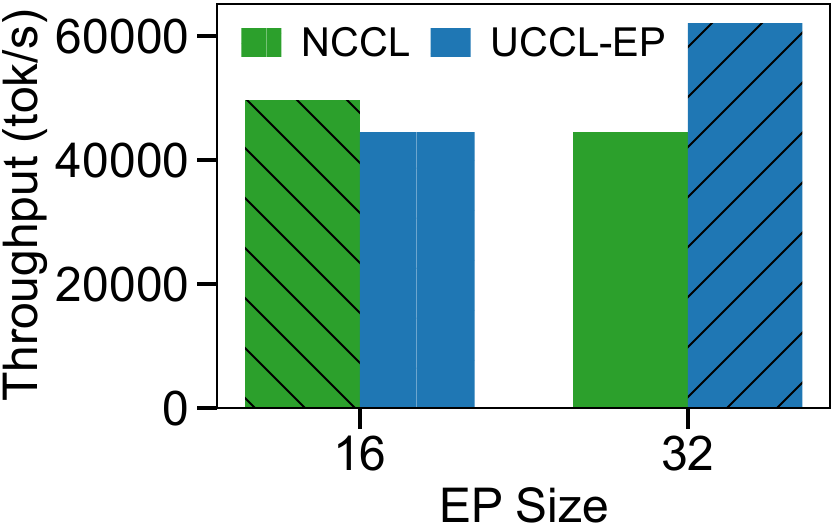}
        \caption{Qwen3 (235B)}
        \label{fig:sglang-qwen3}
    \end{subfigure}
    \vspace{-0.1in}
    \caption{SGLang throughput comparison across two MoE models using DeepEP and NCCL on \pfive. 
    }
    \vspace{-0.1in}
    \label{fig:sglang-comparison}
\end{figure}

\subsubsection{Inference on SGLang with EFA}
\label{ssec:eval-inference}

We evaluate \sysname in SGLang v0.5.3 with EP set to either 16 or 32 on a prefill-heavy workload (input length 4096, output length 5) using \texttt{deepseek-ai/DeepSeek-R1-0528}~\cite{guo2025deepseek} and \texttt{Qwen/Qwen3-235B-A22B-FP8}~\cite{yang2025qwen3}. The results are presented in Figure~\ref{fig:sglang-comparison}. We compare against NCCL, as DeepEP does not run on EFA; at the time of writing, PPLX had not been integrated into open-sourced inference engine. For DeepSeek R1 at EP=16, \sysname reaches an input throughput of 46K~tok/s, about 5\% higher than NCCL. Scaling to EP=32, \sysname further improves to 74K~tok/s input, a \(1.6\times\) prefill speedup over its own EP=16 run. For Qwen3, at EP=32, \sysname reaches 62K~tok/s throughput versus 44K~tok/s for NCCL (about \(40\%\) higher). Overall, \sysname achieves higher throughput at EP=32 and enables larger EP configurations where NCCL either underperforms or cannot run~\cite{github-sglang-issue3491, github-vllm-issue12256} (confirmed with SGLang maintainers). We also observe that CPU utilization increases modestly with \sysname, from an average 8\% CPU utilization to 22\% utilization.

\subsubsection{Training on AMD Primus/Megatron-LM}
\label{ssec:eval-training}

\begin{figure}[!t]
    \centering
        \begin{subfigure}[t]{0.48\linewidth}
        \centering
        \includegraphics[width=\linewidth]{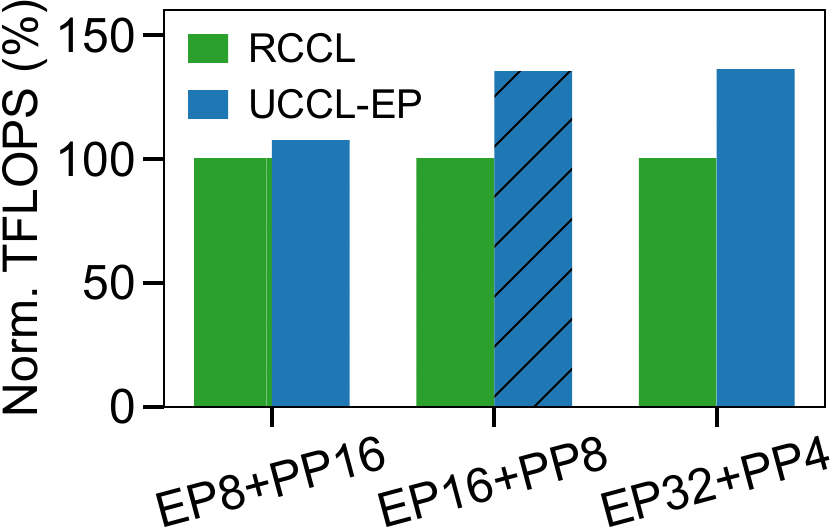}
        \caption{TFLOPS/GPU.}
        \label{fig:training-throughput-tflops}
    \end{subfigure}
    \hfill
    \centering
        \begin{subfigure}[t]{0.48\linewidth}
        \centering
        \includegraphics[width=\linewidth]{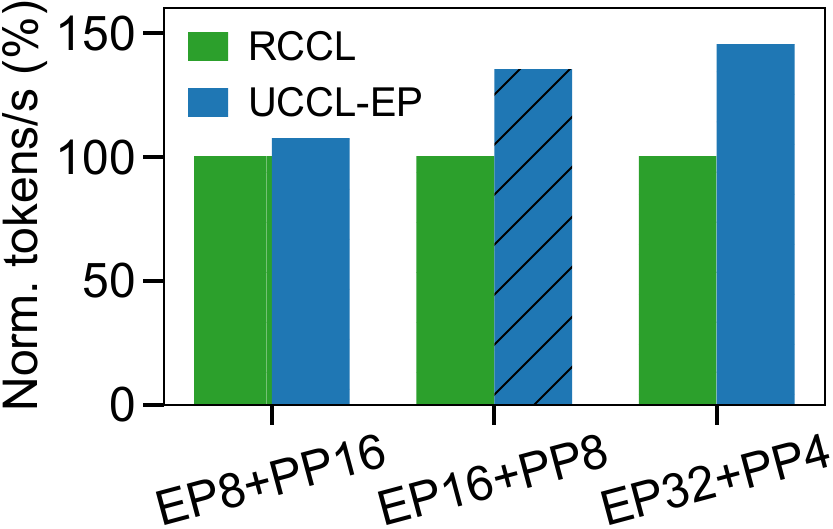}
        \caption{Tokens/s.}
        \label{fig:training-throughput-tokens}
    \end{subfigure}
    \vspace{-0.1in}
    \caption{\amd{Training throughput of DeepSeek-V3 (downscaled to 32 layers and 379B parameters) under AMD Primus/Megatron-LM~\cite{amd_primus}.}}
    \vspace{-0.1in}
    \label{fig:training-throughput}
\end{figure}

Figure~\ref{fig:training-throughput} reports end-to-end Megatron-LM training performance in TFLOPS and tokens per second for DeepSeek V3~\cite{deepseek_v3} over 16 servers. 
Across all models, \sysname matches or exceeds the TFLOPS (by 7-36\%) and throughput achieved by RCCL (by 7-45\%). 
These results show that \sysname leads to performance benefits compared to RCCL for Megatron-LM training on AMD.

\subsection{Design drill-down}
\label{ssec:drill_down}

\begin{figure}[!t]
    \centering
    \includegraphics[width=0.8\linewidth]{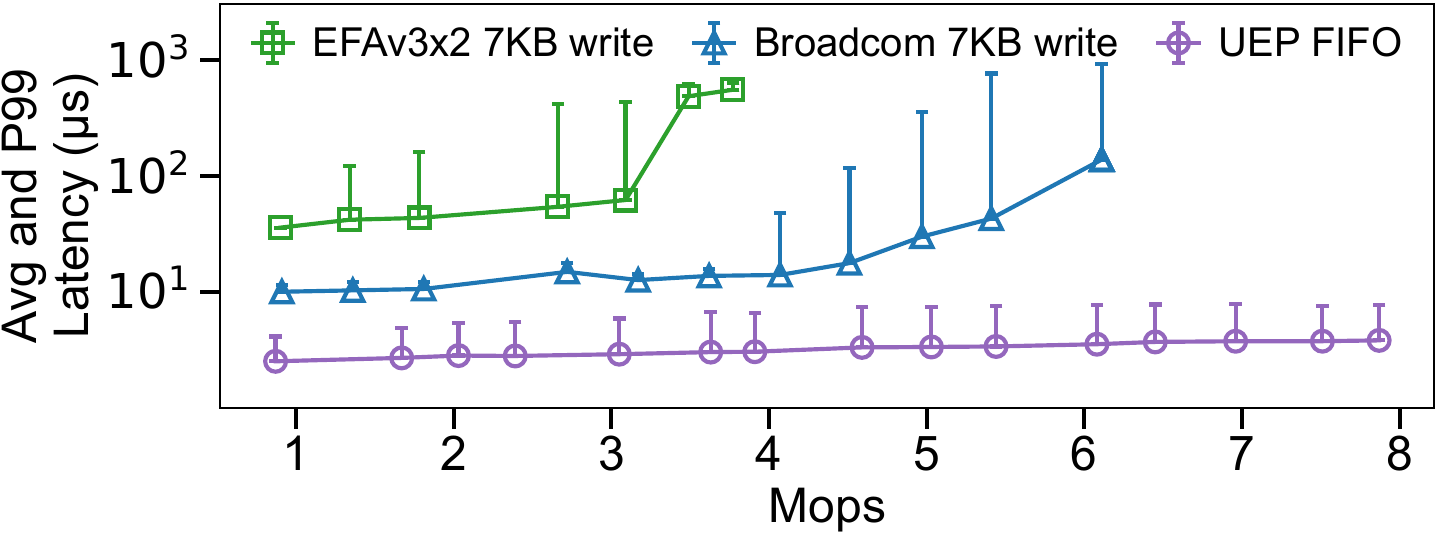}
    \vspace{-0.1in}
    \caption{\sysname FIFO performance. We run 8 FIFOs simultaneously on \pfive and \amdbrc (each for one GPU) and report the first FIFO's result. Note that the y-axis is in log scale.}
    \vspace{-0.1in}
    \label{fig:fifo_latency}
\end{figure}

\paragraph{Stress testing \sysname FIFOs.}
Figure~\ref{fig:fifo_latency} benchmarks how the latency of the FIFO queue compares to network latency with increasing 7KB message throughput. The latency incurred along the FIFO queue is an order of magnitude smaller than the network latency. Furthermore, \sysname FIFO queues are able to scale to large QPS (\eg, 8 Mops), capable of handling modern MoE workloads. 
Note that this benchmark does not batch requests from GPUs to CPUs; we believe that with lightweight batching, it is feasible to handle the next-generation 800G network~\cite{nvidia_800g}.

\begin{figure}[!t]
    \centering
    \begin{subfigure}[t]{0.49\linewidth}
        \centering
        \includegraphics[width=\linewidth]{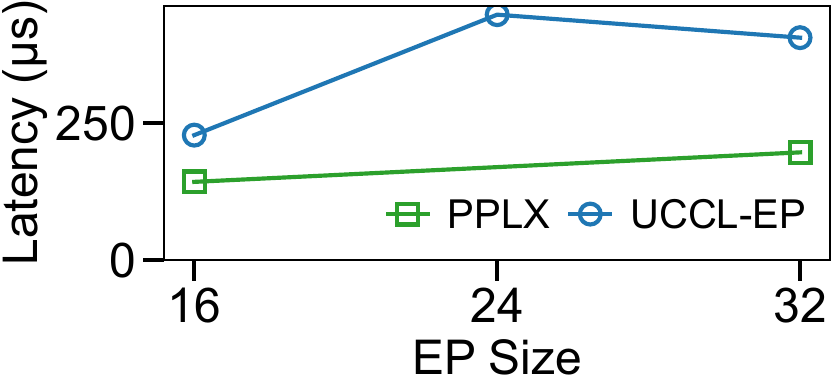}
        \caption{LL}
        \label{fig:ll-sensitivity-ep-degree}
    \end{subfigure}
    \hfill
    \begin{subfigure}[t]{0.49\linewidth}
        \centering
        \includegraphics[width=\linewidth]{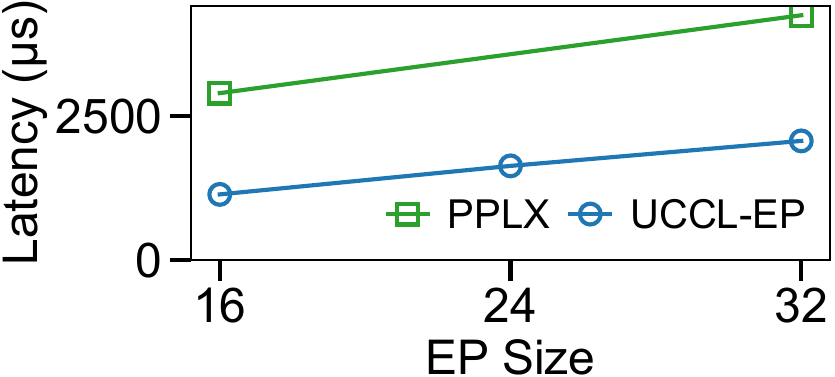}
        \caption{HT}
        \label{fig:ht-sensitivity-ep-degree}
    \end{subfigure}
    \vspace{-0.1in}
    \caption{Sensitivity to EP size. We omit EP24 results for PPLX, as our rented AWS VMs run out of time. 
    }
    \vspace{-0.1in}
    \label{fig:sensitivity-to-ep-degree}
\end{figure}

\paragraph{Varying EP degrees.}
Figure~\ref{fig:sensitivity-to-ep-degree} shows that \sysname achieves better latency compared to PPLX in HT mode, but higher latency in LL mode. This primarily stems from \sysname (extending and therefore similar to DeepEP) LL kernel issuing one token (7KB) per message, and current EFA firmware not being able to process small tokens quickly enough (See Figure~\ref{fig:fifo_latency}. We discuss more in \S\ref{sec:discussion}). 

\begin{figure}[!t]
    \centering
    \begin{subfigure}[t]{0.49\linewidth}
        \centering
        \includegraphics[width=\linewidth]{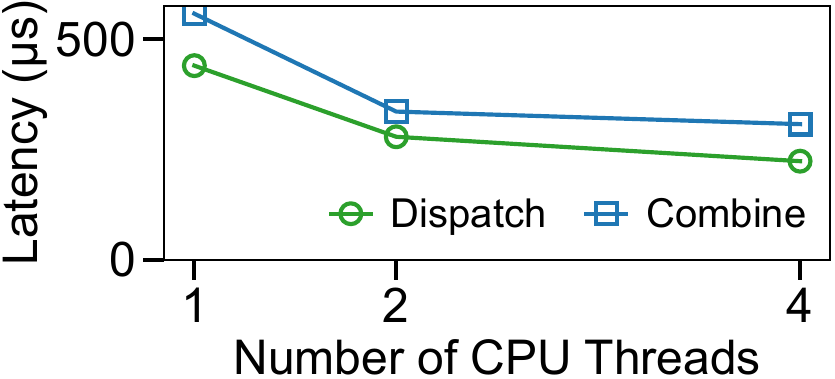}
        \caption{LL}
        \label{fig:ll-sensitivity-cpu-thread}
    \end{subfigure}
    \hfill
    \begin{subfigure}[t]{0.49\linewidth}
        \centering
        \includegraphics[width=\linewidth]{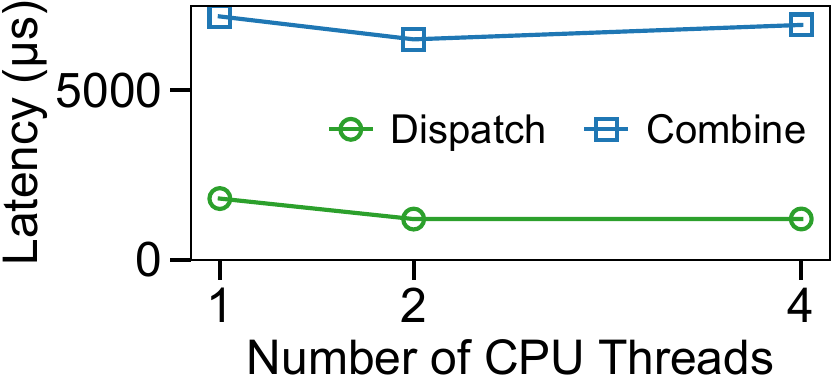}
        \caption{HT}
        \label{fig:ht-sensitivity-cpu-thread}
    \end{subfigure}
    \vspace{-0.1in}
    \caption{Sensitivity to number of CPU threads (\pfive).}
    \vspace{-0.1in}
    \label{fig:sensitivity-to-cpu-threads}
\end{figure}

\paragraph{Varying number of proxy threads.} Figure~\ref{fig:sensitivity-to-cpu-threads} shows that LL and HT kernels suffer from suboptimal performance when the number of CPU threads is 1. The performance significantly improves when we add more CPU threads (\eg, to 4). With more threads per CPU proxy, \sysname is able to use more cores to drive higher-throughput communication. We note that CPU utilization is typically low, as  GPU servers typically have many CPU cores (Table~\ref{tab:eval_testbed}); we observe that even with 4 CPU threads, the CPU utilization only increases modestly. 

\section{Discussion and Future Work}
\label{sec:discussion}

\paragraph{Congestion control with CPU proxy.}
We observe that the number of outstanding requests can have a significant impact on various NICs, affecting tail latency\cite{perplexity2025efa, zhu2025megascale}. This becomes increasingly significant as the number of destinations increases, where having one straggler can significantly slow down dispatch and combine time. 
Throttling or flow control is challenging to implement with IBGDA, as GPU threads typically are not flexible enough for implementing network-level policies. 

Instead, \sysname delegates \textit{control decisions} to the flexible CPU proxy, which could easily support request tracking and pacing. If the outstanding requests become high, the CPU proxy thread temporarily buffers the messages at the sender, so that the messages will not cause an incast at the receiver side. 
The CPU proxy could also bear responsibility for multi‐QP management: because each GPU may be connected to multiple NICs or multiple QPs, the proxy can throttle or shard the outgoing requests across NICs and QPs to avoid congestion~\cite{uccl_collective}, and adapt to NIC‐specific characteristics (\eg, maximum outstanding write requests) without burdening the GPU kernel logic. 
This separation allows exploiting multiple NICs transparently and optimizing throughput while still keeping the GPU's logic simple. 

\paragraph{Elastic EP with CPU proxy.}
Many \amd{existing GPU-initiated communication frameworks assume that RDMA operations succeed and expose limited error-handling and recovery mechanisms to GPU kernels. In practical deployments, however, events such as GPU failures, node additions, or removals can occur and typically require restarting the communication system.} \sysname could achieve elastic EP 
by using the flexible CPU proxy to handle failure and scaling events, transparently masking them from the GPU kernel.

\paragraph{More efficient low-latency kernel.} 
\sysname's low-latency kernel 
could be further optimized by packing tokens in a best-effort manner before sending them out as PPLX does. 
This would particularly benefit AWS EFA NICs. 
We deem these optimizations as orthogonal to \sysname's contribution in a portable expert-parallel communication architecture by decoupling NIC and GPUs.

\paragraph{Supporting AI accelerators} \sysname can be extended to support efficient expert-parallel communication on AI accelerators such as TPUs and AWS Trainium~\cite{aws_trainium_2025}. We believe these are important future directions. 
\section{Related Works}
\label{sec:related-works}

EP communication systems mainly have two categories. 

\paragraph{CPU-initiated.} 
CPU-initiated communication, such as NCCL~\cite{nccl} and RCCL~\cite{rccl} (for AMD GPUs)  relies on the CPU to initiate communication, including constructing and posting RDMA work requests, polling completion queues, and managing QP state transitions. 
Nearly all MoE training and serving frameworks support NCCL/RCCL for MoE communication. 
However, as mentioned in \S\ref{ssec:ep_fine_grained_comm}, they would require either re-packing tokens or transferring small tokens, and do not support token deduplication and hierarchical reduce, both leading to low performance as the number of tokens increases. 
Recent efforts such as UCCL~\cite{uccl_collective} and NCCLX~\cite{ncclx} preserve NCCL/RCCL's CPU-driven design, with various performance optimizations: 
UCCL leverages software packet spraying to utilize multiple network paths and avoid single-path congestion. 
NCCLX introduces topology-aware collective algorithms, zero-copy, and SM-free transfers. 
Both are orthogonal to \sysname. 

\paragraph{GPU-initiated.} 
GPU-initiated communication, by contrast, issues network transfers directly from the GPU, enabling fine-grained token operations like deduplication and hierarchical reduce that are critical to EP communication efficiency. 
DeepEP~\cite{deepep2025}, Mooncake-EP~\cite{mooncake_ep}, 
and ROCm-DeepEP~\cite{ROCmDeepEP} (based on rocSHMEM~\cite{rocshmem}) belong to this category. 
Although providing high performance, they only work for either NVIDIA GPUs or AMD GPUs, and require strict ordering for the underlying NICs. 
NVIDIA's recent HybridEP~\cite{hyridep} atop of DeepEP supports intra-node and multi-node NVLink scenarios with reduced SM usage, while \sysname focuses on inter-node RDMA scenarios. 

MSCCL++~\cite{ShahJLRHJMSCZDMY2025} adopts the GPU-initiated approach to implement NCCL/RCCL collectives such as all-reduce and reduce-scatter, but does not support irregular EP communication. 
The CPU-assisted IBGDA~\cite{ibgda_cpu_assisted} and IBRC~\cite{ibrc} transports from NVSHMEM~\cite{nvidia_nvshmem} could theoretically support non-NVIDIA GPUs via their CPU proxy designs.
But they suffer from lower performance for small token activations due to single-threaded proxies and assume strict ordering for the NICs. 
The EFA transport~\cite{nvshmem_efa} from NVSHMEM supports AWS EFA NICs, but suffers from poor performance~\cite{pplx_garden} by using a single-threaded proxy. 
Compared to them, \sysname supports both heterogeneous GPUs and NICs, enables fine-grained token operations, and provides high performance even with small tokens. 

\section{Conclusion}

Modern MoE workloads require fast and scalable expert-parallel communication, yet existing systems that support token-level GPU-initiated communication like DeepEP,
\amd{remain tightly coupled to specific GPU–NIC combinations and delivery assumptions, limiting portability across increasingly heterogeneous accelerators and networking platforms.} Such an approach results in brittle systems: the communication kernel requires strict delivery semantics from the underlying networking stack, and with limited visibility into error handling, congestion control, and network management. 
This paper introduces \sysname, a portable EP communication system that achieves DeepEP-level performance without relying on specialized GPU–NIC integrations. Our implementation across NVIDIA and AMD GPUs and multiple NIC vendors shows that \sysname outperforms the best existing EP solution by $2.1\times$. 
\sysname is a drop-in replacement for DeepEP applications such as SGLang and AMD Primus/Megatron-LM, improving token throughput by up to 40\% on an NVIDIA+EFA platform, and DeepSeek-V3 training throughput by up to 45\% on a 16-node AMD+Broadcom platform. 
\section*{Acknowledgement}

This work is in part supported by gifts from
from Accenture, AMD, Anyscale, Broadcom, Cisco, Google, IBM, Intel, Intesa Sanpaolo, Lambda, Lightspeed, Mibura, Microsoft, NVIDIA, Samsung SDS, and SAP. 
We additionally acknowledge AWS, in particular Jun Wu, Yida Wang, Brian Barrett, and Nafea Bshara, for their sponsorship and partnership in this research.

\bibliographystyle{abbrv}
\bibliography{biblio.bib}

@misc{deepep2025,
      title={DeepEP: an efficient expert-parallel communication library},
      author={Chenggang Zhao and Shangyan Zhou and Liyue Zhang and Chengqi Deng and Zhean Xu and Yuxuan Liu and Kuai Yu and Jiashi Li and Liang Zhao},
      year={2025},
      publisher = {GitHub},
      howpublished = {\url{https://github.com/deepseek-ai/DeepEP}},
}

@misc{deepseek-ai_EPLB_2025,
  author       = {{deepseek-ai}},
  title        = {EPLB: Expert Parallelism Load Balancer},
  howpublished = {\url{https://github.com/deepseek-ai/EPLB}},
  year         = {2025},
  note         = {GitHub repository, last accessed YYYY-MM-DD}
}

@misc{pplx-kernels,
  title        = {{pplx-kernels}: Perplexity MoE Kernels},
  author       = {Licker, Nandor and Hu, Kevin and Zaytsev, Vladimir and Chen, Lequn},
  year         = {2025},
  publisher    = {GitHub},
  howpublished = {\url{https://github.com/perplexityai/pplx-kernels}},
}

@inproceedings{mao2025skyserve,
  title={Skyserve: Serving ai models across regions and clouds with spot instances},
  author={Mao, Ziming and Xia, Tian and Wu, Zhanghao and Chiang, Wei-Lin and Griggs, Tyler and Bhardwaj, Romil and Yang, Zongheng and Shenker, Scott and Stoica, Ion},
  booktitle={Proceedings of the Twentieth European Conference on Computer Systems},
  pages={159--175},
  year={2025}
}

@misc{aws_trainium_2025,
  author       = {{Amazon Web Services}},
  title        = {AWS Trainium},
  year         = {2025},
  howpublished = {\url{https://aws.amazon.com/ai/machine-learning/trainium/}},
  note         = {Accessed: 2025-12-22}
}

@article{zheng2024sglang,
  title={Sglang: Efficient execution of structured language model programs},
  author={Zheng, Lianmin and Yin, Liangsheng and Xie, Zhiqiang and Sun, Chuyue Livia and Huang, Jeff and Yu, Cody Hao and Cao, Shiyi and Kozyrakis, Christos and Stoica, Ion and Gonzalez, Joseph E and others},
  journal={Advances in neural information processing systems},
  volume={37},
  pages={62557--62583},
  year={2024}
}

@article{griggs2024m,
  title={M$\backslash$'elange: Cost efficient large language model serving by exploiting gpu heterogeneity},
  author={Griggs, Tyler and Liu, Xiaoxuan and Yu, Jiaxiang and Kim, Doyoung and Chiang, Wei-Lin and Cheung, Alvin and Stoica, Ion},
  journal={arXiv preprint arXiv:2404.14527},
  year={2024}
}

@misc{github-vllm-issue12256,
  author       = {DavideHe},
  title        = {[Usage]: deepseek v3 cannot set tensor\_parallel\_size=32 (Issue \#12256)},
  howpublished = {\url{https://github.com/vllm-project/vllm/issues/12256}},
  note         = {GitHub issue on the vllm-project/vllm repository. Accessed: 2025-12-10},
  year         = {2025},
  month        = jan,
  day          = {21}
}

@misc{github-sglang-issue3491,
  author       = {TexasRangers86},
  title        = {[Bug] deepseek-r1 with 4*A100 got error (Issue \#3491)},
  howpublished = {\url{https://github.com/sgl-project/sglang/issues/3491}},
  note         = {GitHub issue on the sgl-project/sglang repository. Accessed: 2025-12-10},
  year         = {2025},
  month        = feb,
  day          = {11}
}

@misc{nvshmem_mlx5_ifc_2025,
  author       = {{NVIDIA/nvshmem contributors}},
  title        = {mlx5\_ifc.h — NVSHMEM MLX5 Transport Interface Header},
  howpublished = {\url{https://github.com/NVIDIA/nvshmem/blob/devel/src/modules/transport/common/mlx5_ifc.h}},
  year         = {2025},
  note         = {Accessed: 2025-12-11},
  organization = {GitHub}
}

@misc{mori_bnxt_re_hsi_2025,
  author       = {{ROCm/mori contributors}},
  title        = {bnxt\_re\_hsi.h — Mori RDMA Provider BNXT HSI Header},
  howpublished = {\url{https://github.com/ROCm/mori/blob/main/include/mori/core/transport/rdma/providers/bnxt/bnxt_re_hsi.h}},
  year         = {2025},
  note         = {Accessed: 2025-12-11},
  organization = {GitHub}
}

@misc{mooncake_mlx5_ifc_2025,
  author       = {{kvcache-ai/Mooncake contributors}},
  title        = {mlx5\_ifc.h — Mooncake IBGDA MLX5 Interface Header},
  howpublished = {\url{https://github.com/kvcache-ai/Mooncake/blob/main/mooncake-ep/include/mooncake_ibgda/mlx5_ifc.h}},
  year         = {2025},
  note         = {Accessed: 2025-12-11},
  organization = {GitHub}
}

@article{Langer2024_NVSHMEM3,
  title        = {Enhancing Application Portability and Compatibility across New Platforms Using NVIDIA Magnum IO NVSHMEM 3.0},
  author       = {Akhil Langer and Seth Howell and Arnav Goel and Pak Markthub and Harry Petty and Fred Oh},
  year         = {2024},
  month        = sep # " 06",
  journal      = {NVIDIA Technical Blog},
  url          = {https://developer.nvidia.com/blog/enhancing-application-portability-and-compatibility-across-new-platforms-using-nvidia-magnum-io-nvshmem-3-0/},
}

@article{yang2025qwen3,
  title={Qwen3 technical report},
  author={Yang, An and Li, Anfeng and Yang, Baosong and Zhang, Beichen and Hui, Binyuan and Zheng, Bo and Yu, Bowen and Gao, Chang and Huang, Chengen and Lv, Chenxu and others},
  journal={arXiv preprint arXiv:2505.09388},
  year={2025}
}

@article{guo2025deepseek,
  title={Deepseek-r1: Incentivizing reasoning capability in llms via reinforcement learning},
  author={Guo, Daya and Yang, Dejian and Zhang, Haowei and Song, Junxiao and Zhang, Ruoyu and Xu, Runxin and Zhu, Qihao and Ma, Shirong and Wang, Peiyi and Bi, Xiao and others},
  journal={arXiv preprint arXiv:2501.12948},
  year={2025}
}

@misc{nvidia_gh200_grace_hopper_superchip,
  author       = {{NVIDIA Corporation}},
  title        = {NVIDIA GH200 Grace Hopper Superchip},
  year         = {2025},
  howpublished = {\url{https://www.nvidia.com/en-us/data-center/grace-hopper-superchip/}},
  note         = {Accessed: 2025-12-08}
}

@misc{ROCmDeepEP,
  author       = {{ROCm}},
  title        = {DeepEP: a high-performance expert-parallel communication library},
  year         = {2025},
  howpublished = {\url{https://github.com/ROCm/DeepEP}},
  note         = {Accessed: 2025-12-05}
}

@article{jiang2025hexgen,
  title={Hexgen-2: Disaggregated generative inference of llms in heterogeneous environment},
  author={Jiang, Youhe and Yan, Ran and Yuan, Binhang},
  journal={arXiv preprint arXiv:2502.07903},
  year={2025}
}

@article{zhang2025cauchy,
  title={Cauchy: A Cost-Efficient LLM Serving System through Adaptive Heterogeneous Deployment},
  author={Zhang, Yihui and Shen, Han and Yang, Renyu and Tian, Di and Luo, Yuxi and Zhang, Menghao and Li, Li and Hu, Chunming and Wo, Tianyu and Song, Chengru and others},
  year={2025}
}

@article{li2025taming,
  title={Taming the Chaos: Coordinated Autoscaling for Heterogeneous and Disaggregated LLM Inference},
  author={Li, Rongzhi and Du, Ruogu and Chu, Zefang and Zhao, Sida and Han, Chunlei and Shi, Zuocheng and Shao, Yiwen and Han, Huanle and Huang, Long and Liu, Zherui and others},
  journal={arXiv preprint arXiv:2508.19559},
  year={2025}
}

@inproceedings{mei2025helix,
  title={Helix: Serving large language models over heterogeneous gpus and network via max-flow},
  author={Mei, Yixuan and Zhuang, Yonghao and Miao, Xupeng and Yang, Juncheng and Jia, Zhihao and Vinayak, Rashmi},
  booktitle={Proceedings of the 30th ACM International Conference on Architectural Support for Programming Languages and Operating Systems, Volume 1},
  pages={586--602},
  year={2025}
}

@article{jaiswal2025serving,
  title={Serving models, fast and slow: optimizing heterogeneous llm inferencing workloads at scale},
  author={Jaiswal, Shashwat and Jain, Kunal and Simmhan, Yogesh and Parayil, Anjaly and Mallick, Ankur and Wang, Rujia and St Amant, Renee and Bansal, Chetan and R{\"u}hle, Victor and Kulkarni, Anoop and others},
  journal={arXiv e-prints},
  pages={arXiv--2502},
  year={2025}
}

@misc{ShahJLRHJMSCZDMY2025,
  title        = {MSCCL++: Rethinking GPU Communication Abstractions for Cutting-edge AI Applications},
  author       = {Aashaka Shah and Abhinav Jangda and Binyang Li and Caio Rocha and Changho Hwang and Jithin Jose and Madan Musuvathi and Olli Saarikivi and Peng Cheng and Qinghua Zhou and Roshan Dathathri and Saeed Maleki and Ziyue Yang},
  year         = {2025},
  eprint       = {2504.09014},
  archivePrefix= {arXiv},
  primaryClass = {cs.DC},
  url          = {https://arxiv.org/abs/2504.09014},
}

@article{deepseek_v3,
  title={{DeepSeek-V3 Technical Report}},
  author={Liu, Aixin and Feng, Bei and Xue, Bing and Wang, Bingxuan and Wu, Bochao and Lu, Chengda and Zhao, Chenggang and Deng, Chengqi and Zhang, Chenyu and Ruan, Chong and others},
  journal={arXiv preprint arXiv:2412.19437},
  year={2024}
}

@misc{nccl,
    title = {{NVIDIA Collective Communications Library (NCCL)}},
    author = {{NVIDIA Corporation}},
    year = {2025},
    howpublished = {\url{https://github.com/NVIDIA/nccl}}
}

@misc{rccl,
    title = {{ROCm Communication Collectives Library (RCCL)}},
    author = {{AMD}},
    year = {2025},
    howpublished = {\url{https://github.com/ROCm/rccl}}
}

@misc{efa,
  title={{Elastic Fabric Adapter}},
  author={Amazon Web Services},
  year = {2025},
  howpublished = {\url{https://aws.amazon.com/hpc/efa/}}
}

@misc{ibgda,
  title={{Using the NVSHMEM InfiniBand GPUDirect Async Transport}},
  author = {{NVIDIA Corporation}},
  year = {2025},
  howpublished = {\url{https://docs.nvidia.com/nvshmem/api/using.html##using-the-nvshmem-infiniband-gpudirect-async-transport}}
}

@misc{ibgda_cpu_assisted,
  title={{CPU-assisted InfiniBand GPU Direct Async}},
  author = {{NVIDIA Corporation}},
  year = {2024},
  howpublished = {\url{https://developer.nvidia.com/blog/enhancing-application-portability-and-compatibility-across-new-platforms-using-nvidia-magnum-io-nvshmem-3-0/##cpu-assisted_infiniband_gpu_direct_async\%C2\%A0}}
}

@inproceedings{deepseek_isca,
  title={Insights into deepseek-v3: Scaling challenges and reflections on hardware for ai architectures},
  author={Zhao, Chenggang and Deng, Chengqi and Ruan, Chong and Dai, Damai and Gao, Huazuo and Li, Jiashi and Zhang, Liyue and Huang, Panpan and Zhou, Shangyan and Ma, Shirong and others},
  booktitle={Proceedings of the 52nd Annual International Symposium on Computer Architecture},
  pages={1731--1745},
  year={2025}
}

@inproceedings{characterization_llm_cluster,
  title={Characterization of large language model development in the datacenter},
  author={Hu, Qinghao and Ye, Zhisheng and Wang, Zerui and Wang, Guoteng and Zhang, Meng and Chen, Qiaoling and Sun, Peng and Lin, Dahua and Wang, Xiaolin and Luo, Yingwei and others},
  booktitle={Proceedings of USENIX NSDI},
  pages={709--729},
  year={2024}
}

@inproceedings{byteps,
  title={A unified architecture for accelerating distributed $\{$DNN$\}$ training in heterogeneous $\{$GPU/CPU$\}$ clusters},
  author={Jiang, Yimin and Zhu, Yibo and Lan, Chang and Yi, Bairen and Cui, Yong and Guo, Chuanxiong},
  booktitle={Proceedings of USENIX OSDI},
  pages={463--479},
  year={2020}
}

@misc{megatronlm,
  title={{Megatron-LM \& Megatron-Core}},
  author = {{NVIDIA Corporation}},
  year = {2025},
  howpublished = {\url{https://github.com/NVIDIA/Megatron-LM}}
}

@article{deepseek_v3_2,
  title   = {DeepSeek-V3.2: Pushing the Frontier of Open Large Language Models},
  author  = {DeepSeek-AI and Aixin Liu and Aoxue Mei and Bangcai Lin and Bing Xue and Bingxuan Wang and Bingzheng Xu and Bochao Wu and Bowei Zhang and Chaofan Lin and Chen Dong and Chengda Lu and Chenggang Zhao and Chengqi Deng and Chenhao Xu and Chong Ruan and Damai Dai and Daya Guo and Dejian Yang and ... and Zihui Gu and Zijia Zhu and Zilin Li and Zipeng Zhang},
  journal = {arXiv},
  volume  = {abs/2512.02556},
  year    = {2025},
  url     = {https://arxiv.org/abs/2512.02556}
}

@misc{gemini3pro,
  title        = {Gemini 3 Pro},
  author       = {Google DeepMind / Google},
  year         = {2025},
  note         = {Released November 18, 2025},
  url          = {https://cloud.google.com/vertex-ai/docs/generative-ai/models}
}

@misc{ibta_spec,
  title        = {InfiniBand Trade Association — IBTA Specification Portal},
  organization = {InfiniBand Trade Association},
  url          = {https://www.infinibandta.org/ibta-specification/},
  note         = {Accessed 2025},
}

@misc{sglang_deepep,
  title        = {Deploying DeepSeek with PD Disaggregation and Large-Scale Expert Parallelism on 96 H100 GPUs},
  author       = {{The SGLang Team}},
  year         = {2025},
  howpublished = {\url{https://lmsys.org/blog/2025-05-05-large-scale-ep/}},
  note         = {Accessed 2025},
}

@misc{megatron_lm_deepep,
  author       = {NVIDIA/Megatron-LM Contributors},
  title        = {Benchmarking DeepEP Guide \#1721},
  howpublished = {\url{https://github.com/nvidia/megatron-lm/issues/1721}},
  year         = {2025},
  note         = {GitHub issue, accessed 2025},
}

@misc{vllm_deepep,
  title        = {Expert Parallel Deployment},
  author       = {{vLLM Documentation Team}},
  howpublished = {\url{https://docs.vllm.ai/en/latest/serving/expert_parallel_deployment/}},
  year         = {2025},
  note         = {Accessed 2025},
}

@article{aws_srd,
  title={{A Cloud-Optimized Transport Protocol for Elastic and Scalable HPC}},
  author={Shalev, Leah and Ayoub, Hani and Bshara, Nafea and Sabbag, Erez},
  journal={IEEE micro},
  volume={40},
  number={6},
  pages={67--73},
  year={2020},
  publisher={IEEE}
}

@misc{cx7,
  title={{ConnectX-7 400G Adapters}},
  author={{NVIDIA Corporation}},
  year = {2024},
  howpublished = {\url{https://resources.nvidia.com/en-us-accelerated-networking-resource-library/connectx-7-datasheet}}
}

@article{c4_alibaba,
  title={{Boosting Large-Scale Parallel Training Efficiency with C4: A Communication-Driven Approach}},
  author={Dong, Jianbo and Luo, Bin and Zhang, Jun and Zhang, Pengcheng and Feng, Fei and Zhu, Yikai and Liu, Ang and Chen, Zian and Shi, Yi and Jiao, Hairong and others},
  journal={arXiv preprint arXiv:2406.04594},
  year={2024}
}

@misc{tma_engine,
  author       = {{Andersch, Michael and Palmer, Greg and Krashinsky, Ronny and Stam, Nick and Mehta, Vishal and Brito, Gonzalo and Ramaswamy, Sridhar}},
  title        = {NVIDIA Hopper Architecture In-Depth},
  year         = {2022},
  month        = mar,
  howpublished = {\url{https://developer.nvidia.com/blog/nvidia-hopper-architecture-in-depth/}},
  note         = {NVIDIA Developer Blog; Accessed: 2025-12-06}
}

@article{du2022glam,
  title        = {Efficient Scaling of Language Models with Mixture‐of‐Experts},
  author       = {Du, Nan and Huang, Yanping and Dai, Andrew M. and Tong, Simon and Lepikhin, Dmitry and Xu, Yuanzhong and Krikun, Maxim and Zhou, Yanqi and Wei Yu, Adams and Firat, Orhan and Zoph, Barret and Dixon, Lucas and Chen, Zhifeng and Cui, Claire},
  journal      = {arXiv preprint arXiv:2112.06905},
  year         = {2022},
  url          = {https://arxiv.org/pdf/2112.06905},
  note         = {GLaM (Generalist Language Model) introduced by Google as a sparse MoE model} 
}

@article{jiang2024mixtral,
  title        = {Mixtral of Experts},
  author       = {Jiang, Albert Q. and Sablayrolles, Alexandre and Roux, Antoine and Mensch, Arthur and Savary, Blanche and Bamford, Chris and Chaplot, Devendra S. and de las Casas, Diego and Bou Hanna, Emma and Bressand, Florian and Lengyel, Gianna and Lample, Guillaume and Lavaud, Lélio R. and Saulnier, Lucile and Lachaux, Marie-Anne and Stock, Pierre and Subramanian, Sandeep},
  journal      = {arXiv preprint arXiv:2401.04088},
  year         = {2024},
  url          = {https://arxiv.org/abs/2401.04088},
  note         = {Mixtral 8×7B sparse Mixture-of-Experts model by Mistral AI}
}

@inproceedings{dai2024deepseekmoe,
  title        = {DeepSeekMoE: Towards Ultimate Expert Specialization in Mixture-of-Experts Language Models},
  author       = {Dai, Damai and Deng, Chengqi and Zhao, Chenggang and Xu, R.X. and Gao, Huazuo and Chen, Deli and Li, Jiashi and Zeng, Wangding and Yu, Xingkai and Wu, Y. and Xie, Zhenda and Li, Y.K. and Huang, Panpan and Luo, Fuli and Ruan, Chong and Sui, Zhifang and Liang, Wenfeng},
  booktitle    = {Proceedings of the 62nd Annual Meeting of the Association for Computational Linguistics (Long Papers)},
  pages        = {1280--1297},
  year         = {2024},
  organization = {Association for Computational Linguistics},
  url          = {https://aclanthology.org/2024.acl-long.70},
  note         = {DeepSeekMoE model paper}
}

@article{muennighoff2024olmoe,
  title        = {OLMoE: Open Mixture-of-Experts Language Models},
  author       = {Muennighoff, Niklas and Soldaini, Luca and Groeneveld, Dirk and others},
  journal      = {arXiv preprint arXiv:2409.02060},
  year         = {2024},
  url          = {https://arxiv.org/html/2409.02060},
  note         = {Open Mixture-of-Experts models by the Allen Institute}
}

@misc{meta_llama4_2025,
  author       = {{Meta AI}},
  title        = {Introducing LLaMA 4: Multimodal Intelligence with Mixture-of-Experts},
  year         = {2025},
  howpublished = {\url{https://ai.meta.com/blog/llama-4-multimodal-intelligence/}},
  note         = {Meta’s official announcement of LLaMA 4 Scout and Maverick}
}

@manual{libibverbs,
  title        = {libibverbs: RDMA Userspace Verbs API},
  author       = {{RDMA Consortium}},
  organization = {Linux RDMA Project},
  year         = {2024},
  url          = {https://github.com/linux-rdma/rdma-core},
  note         = {Accessed: 2025-12-06}
}

@misc{pytorch_symmetric_memory,
  author       = {{PyTorch}},
  title        = {Symmetric Memory},
  howpublished = {\url{https://docs.pytorch.org/docs/stable/symmetric_memory.html}},
  note         = {Accessed: 2025-12-06},
  year         = {2025}
}

@misc{nvidia_nvshmem,
  author       = {{NVIDIA}},
  title        = {NVSHMEM},
  howpublished = {\url{https://developer.nvidia.com/nvshmem}},
  note         = {Accessed: 2025-12-06},
  year         = {2025}
}

@article{uccl_collective,
  title={An Extensible Software Transport Layer for GPU Networking},
  author={Zhou, Yang and Chen, Zhongjie and Mao, Ziming and Lao, ChonLam and Yang, Shuo and Kannan, Pravein Govindan and Gao, Jiaqi and Zhao, Yilong and Wu, Yongji and You, Kaichao and others},
  journal={arXiv preprint arXiv:2504.17307},
  year={2025}
}

@article{ncclx,
  title={Collective communication for 100k+ gpus},
  author={Si, Min and Balaji, Pavan and Chen, Yongzhou and Chu, Ching-Hsiang and Gangidi, Adi and Hasan, Saif and Iyengar, Subodh and Johnson, Dan and Liu, Bingzhe and Ren, Jingliang and others},
  journal={arXiv preprint arXiv:2510.20171},
  year={2025}
}

@article{pplx_garden,
  title={RDMA Point-to-Point Communication for LLM Systems},
  author={Licker, Nandor and Hu, Kevin and Zaytsev, Vladimir and Chen, Lequn},
  journal={arXiv preprint arXiv:2510.27656},
  year={2025}
}

@online{ibrc,
  author       = {Pak Markthub and Jim Dinan and Sreeram Potluri and Seth Howell},
  title        = {Improving Network Performance of HPC Systems Using NVIDIA Magnum IO, NVSHMEM and GPUDirect Async},
  year         = {2022},
  month        = {Nov 22},
  url          = {https://developer.nvidia.com/blog/improving-network-performance-of-hpc-systems-using-nvidia-magnum-io-nvshmem-and-gpudirect-async/},
  organization = {NVIDIA Developer Blog}
}

@misc{hyridep,
  author       = {{NVIDIA Corporation}},
  title        = {HybridEP for High-Performance Intra-Node Token Dispatching},
  howpublished = {\url{https://github.com/deepseek-ai/DeepEP/tree/hybrid-ep}},
  note         = {Technical documentation published within the DeepEP repository},
  year         = {2024}
}

@misc{rocshmem,
  author       = {AMD ROCm Team},
  title        = {{rocSHMEM}: GPU-Centric OpenSHMEM Runtime for AMD ROCm},
  year         = {2025},
  howpublished = {\url{https://github.com/ROCm/rocSHMEM}},
  note         = {Accessed: 2025-12-07}
}

@misc{nvshmem_efa,
  author       = {NVIDIA},
  title        = {NVSHMEM: Libfabric Transport Backend (EFA Support)},
  howpublished = {\url{https://github.com/NVIDIA/nvshmem/blob/9cc869bc28e565e6944c4ddf76976ada4a1ebbf7/src/modules/transport/libfabric/libfabric.h#L192}},
  note         = {Implements NVSHMEM support for AWS EFA via the Libfabric transport layer. Commit: 9cc869bc28e565e6944c4ddf76976ada4a1ebbf7},
  year         = {2025}
}

@misc{gptoss,
  title        = {Introducing gpt-oss: OpenAI's Open-Weight Reasoning Models},
  author       = {OpenAI},
  year         = {2025},
  howpublished = {\url{https://openai.com/index/introducing-gpt-oss/}},
  note         = {Accessed: 2025-12-08}
}

@misc{llama4,
  title        = {Introducing LLaMA 4: Advancing Multimodal Intelligence},
  author       = {Meta AI},
  year         = {2025},
  howpublished = {\url{https://ai.meta.com/blog/llama-4-multimodal-intelligence/}},
  note         = {Accessed: 2025-12-08}
}

@misc{amd_primus,
    title = {{AMD Primus training framework for large-scale foundation model training and inference on AMD GPUs.}},
    author={AMD-AGI/Primus},
    howpublished = {\url{https://github.com/AMD-AGI/Primus}}
}

@inproceedings{meta-training,
author = {Gangidi, Adithya and Miao, Rui and Zheng, Shengbao and Bondu, Sai Jayesh and Goes, Guilherme and Morsy, Hany and Puri, Rohit and Riftadi, Mohammad and Shetty, Ashmitha Jeevaraj and Yang, Jingyi and Zhang, Shuqiang and Fernandez, Mikel Jimenez and Gandham, Shashidhar and Zeng, Hongyi},
title = {RDMA over Ethernet for Distributed Training at Meta Scale},
year = {2024},
isbn = {9798400706141},
publisher = {Association for Computing Machinery},
address = {New York, NY, USA},
url = {https://doi.org/10.1145/3651890.3672233},
doi = {10.1145/3651890.3672233},
booktitle = {Proceedings of the ACM SIGCOMM 2024 Conference},
pages = {57–70},
numpages = {14},
location = {Sydney, NSW, Australia},
series = {ACM SIGCOMM '24}
}

@misc{deepep_ht,
  author       = {deepseek-ai},
  title        = {{DeepEP}: {\texttt{internode.cu}} at commit \texttt{b57e5e212ab75350f53c72064333e4fe1076b1da}},
  howpublished = {\url{https://github.com/deepseek-ai/DeepEP/blob/b57e5e212ab75350f53c72064333e4fe1076b1da/csrc/kernels/internode.cu\#L1741}},
  year         = {2023},
  note         = {Accessed: 2025-12-09},
}

@misc{nvidia_800g,
  title={{NVIDIA Ethernet SuperNICs: Next-generation networking for the next wave of AI.}},
  author={NVIDIA Corporation},
  year = {2025},
  howpublished = {\url{https://www.nvidia.com/en-us/networking/products/ethernet/supernic/}}
}

@article{gale2023megablocks,
  title={Megablocks: Efficient sparse training with mixture-of-experts},
  author={Gale, Trevor and Narayanan, Deepak and Young, Cliff and Zaharia, Matei},
  journal={Proceedings of Machine Learning and Systems},
  volume={5},
  pages={288--304},
  year={2023}
}

@inproceedings{zhu2025megascale,
  title={Megascale-infer: Efficient mixture-of-experts model serving with disaggregated expert parallelism},
  author={Zhu, Ruidong and Jiang, Ziheng and Jin, Chao and Wu, Peng and Stuardo, Cesar A and Wang, Dongyang and Zhang, Xinlei and Zhou, Huaping and Wei, Haoran and Cheng, Yang and others},
  booktitle={Proceedings of the ACM SIGCOMM 2025 Conference},
  pages={592--608},
  year={2025}
}

@misc{perplexity2025efa,
  title={Enabling Trillion-Parameter Models on AWS EFA},
  author={{Perplexity AI}},
  year={2025},
  month={November},
  howpublished={\url{https://research.perplexity.ai/articles/enabling-trillion-parameter-models-on-aws-efa}},
  note={Accessed: 2025-12-10}
}

@misc{mooncake_ep,
  author       = {kvcache-ai},
  title        = {Mooncake-EP: Expert-Parallel Extension of Mooncake},
  howpublished = {\url{https://github.com/kvcache-ai/Mooncake/tree/main/mooncake-ep}},
  note         = {Accessed: 2025-12-11},
  year         = {2025}
}

\appendix


\end{document}